\documentclass[showpacs]{revtex4}

\usepackage{graphics}
\usepackage{amssymb}
\usepackage{amsmath}
\begin{document}
\title{Three-dimensional black holes via Noether symmetries}

\author{Ugur Camci} \email{ucamci@rwu.edu,ugurcamci@gmail.com}

\bigskip

\affiliation{Department of Chemistry and Physics, Roger Williams University, One Old Ferry Road, Bristol, RI 02809, USA\label{addr1} }


\bigskip

\date{\today}


\begin{abstract}
We investigate the Noether symmetries of the Lagrangian for the stationary rotating BTZ-type three-dimensional spacetimes in $f(R)$ theory of gravity. A detailed analysis of Noether symmetries of (2+1)-dimensional rotating BTZ-type black hole spacetime model is presented. Applying the Noether symmetry approach, the first integrals (constants of motion) for each of Noether symmetries are obtained to look for the exact solutions. After solving the first integral equations depending on the form of the function $f(R)$, we derived some new (2+1)-dimensional rotating BTZ-type black hole solutions. We discussed the physical implications of the derived exact solutions. The thermodynamical properties of the obtained BTZ-type black hole solutions are analyzed by making use of the mass $M$ and the angular momentum $J$ in terms of $r_{\pm}$, where $r_+$ is the event horizon and $r_{-}$ is the inner horizon. Further, it is shown that thermodynamic quantities obey the first law, and the Smarr-like formulas of the solutions we found are obtained.
\end{abstract}


\maketitle
%
\section{Introduction}
\label{intro}

Due to the fact that the simplest theory of gravity possessing some nice properties is in three dimensions, a black hole in three dimensional gravity theory can be a perfect toy model to study its properties. The mathematical complexity of general relativity is significantly softened in order to get further insights on the fundamental nature of gravitation in (2+1)-dimension. The (2+1)-dimensional black hole solutions inherent all typical characteristics that can be found in (3+1) or higher dimensional black holes such as horizon(s), black hole thermodynamics and Hawking radiation. Thermodynamical properties of low dimensional black holes is ill-defined because of a few degrees of freedom. The (2+1)-dimensional vacuum solution in three-dimensional gravity theory is necessarily flat if the cosmological constant is zero, which means that there is no black hole solution of three-dimensional gravity without a cosmological constant \cite{ida2000}.

In (2+1)-dimensional gravity with a negative cosmological constant, Ba\~{n}ados, Teitelboim and Zanelli (BTZ) has found a black hole solution \cite{btz}, which is commonly called as BTZ black hole and describes an asymptotically anti-de Sitter rotating black hole. The prominent feature of this black hole model lies in its simplicity of its construction. Also, it is shown that this black hole solution can have an arbitrarily high entropy \cite{btz}, and plays a good role in the understanding of thermodynamics of black holes \cite{carlip95, L08, akbar07, akbar11}.  Later on, the vacuum BTZ solution is enlarged to include an electric charge $q$, with similar nice properties \cite{carlip95}. Afterwards, Einstein-Maxwell \cite{mtz2000} and Einstein-Maxwell-dilaton \cite{cm94} extensions were also found. Furthermore, BTZ-type solutions have been studied in context of $f(R)$ gravity \cite{fR}, dilaton gravity \cite{dilaton}, teleparallel and $f(T)$ gravities \cite{fT}, and noncommutative geometries \cite{noncom}.
Among the problems studied in the (2+1)-dimensional context so far, it is also mentioned the magnetic counterparts of the electrically charged BTZ solution \cite{cbtz, halilsoy}, gravitational collapse \cite{viqar}, geometric and thermodynamic features of several non-linear models \cite{nl-sigma}, wormholes \cite{wormhole}, or BTZ-like solutions out of the coupling to scalar fields \cite{scalar}.

Noether symmetries are associated with differential equations possessing a Lagrangian, and they describe physical features of differential equations in terms of conservation laws admitted by them \cite{ibrahim}. The Noether symmetry approach without a gauge term (strict Noether symmetry) is a kind of symmetry in which the Lie derivative of the Lagrangian that arise from the metric of interest dragging along a vector field $\bf{X}$ vanishes,i.e., $\pounds_{\bf X} \mathcal{L} = 0$ \cite{camci1, capo08, vakili08, camci2,darabi2013}. This approach has also been used to obtain $f(R)$ gravity models respecting the Noether symmetry. Darabi et. al. \cite{darabi2013} have used the Noether symmetry without a gauge term to produce (2 + 1)-dimensional black hole solutions in $f(R)$ gravity. The Noether symmetry approach with a gauge term is the generalization of strict Noether symmetry as the existence of some extra symmetries is expected \cite{tp2011a,camci2012,yusuf2015,camci2016,Bahamonde:2016}. This approach will be exhaustively considered in the following section. In many of the extended theories of gravity, the dynamical Lagrangian involves any arbitrary functions (the function of $f(R)$ in this paper or the potential or the coupling functions) which are unknown quantities. Using the Noether symmetry approach, the form of the unknown functions in the Lagrangian may be determined. We also note that Noether symmetry with a gauge term is a physical criterion which allows one to find $f(R)$ gravity models compatible with this symmetry.

In this work, we consider stationary rotating BTZ-type three-dimensional spacetimes. Therefore, the affine parameter $\tau$ will be the radial coordinate $r$. Through this paper, we aim to find Noether symmetries of the dynamical Lagrangian $\mathcal{L}$ for the $(2+1)$-dimensional rotating black hole spacetimes in $f(R)$ gravity.
The action for $f(R)$ theory of gravity in three-dimensions is of the form
\begin{equation}
\mathcal{S} = \int d^{3} x \sqrt{-g} f(R) + \mathcal{S}_m ,
\end{equation}
where $g= \det (g_{\mu\nu})$ and $\mathcal{S}_m$ is the matter action. By employing the variational principle with respect to the metric tensor it gives rise to the field equations as
\begin{equation}
f_{R} R_{\mu \nu} - \frac{1}{2} g_{\mu \nu} f(R)  - \nabla_{\mu} \nabla_{\nu} f_R + g_{\mu \nu} \square f_R  = -\frac{1}{2} T^m_{\mu \nu} \, , \label{feqs}
\end{equation}
where $f_R = d f(R) / dR$, $\square = \nabla^{\mu} \nabla_{\mu}$ is the Laplace-Beltrami operator, and $T^m_{\mu \nu} = - \frac{2}{\sqrt{-g}} \frac{\partial S_m}{\partial g^{\mu \nu}}$. The trace of the field equation \eqref{feqs} implies
\begin{eqnarray}
& & R f_{R} + 2 \square f_R  -\frac{3}{2} f = -\frac{T^m}{2}  \, , \label{trace}
\end{eqnarray}
in which $T^m = T^{m \, \mu}_{\,\, \mu}$. Hereafter, we assume $T^m = 0$. In order to emphasize the physical significance after obtaining new solution(s), one can recast the field equations \eqref{feqs} to an appropriate form by introducing a curvature stress-energy tensor and defining effective density $\rho_{_{eff}}$ and pressure $p_{_{eff}}$ as given in section \ref{sec:phys}. Then, we suggest an effective equation of state in the form $p_{_{eff}} = w_{_{eff}} \rho_{_{eff}}$ to study the properties of the obtained solutions, where $w_{_{eff}}$ is the equation of state parameter which may be a constant or a function of $r$. Furthermore, it would be useful to derive the thermodynamical quantities such as temperature and entropy to show that the first law of thermodynamics is satisfied for the solutions we found. We also calculate the Smarr-like mass formulas of our solutions.

The rest of the paper is organized as follows. In the following section, for the dynamical Lagrangian, we present the Noether symmetry approach with a gauge term in an arbitrary dimensional configuration space, where the Lagrangian includes a velocity term. In section \ref{sec:2}, we will focus on the most general (2+1)-dimensional rotating BTZ-type black hole spacetime in the context of Noether symmetry approach. In section \ref{sec:3}, we will use the obtained first integrals related with the Noether symmetries of section \ref{sec:2} to find exact solutions for  $f(R)$ gravity.  The section \ref{sec:phys} is devoted to study physical properties of the solutions shown in section \ref{sec:3}. Conclusions are presented in last section \ref{sec:conc}.

\section{Noether symmetry approach}
\label{sec:1}

The models for gravity theories are expressed in terms of the configuration space variables which are usually the metric coefficients, matter fields, scalar fields, etc. Therefore, the corresponding configuration space of the model is a $d-$ dimensional Riemannian manifold with coordinates $q^i, \, i=1,2,\ldots, d$, in which it is constructed a point-like Lagrangian to produce the dynamics of the model.

The equations of gravity theories can be derived both from the field equations or deduced by a Lagrangian function $\mathcal{L} (\tau, q^i, {q'}^i)$ of the system related to the action $\mathcal{S} = \int{ \mathcal{L} d\tau}$. Here the prime represents the derivative with respect to an affine parameter $\tau$ which is the cosmic time $t$ in most of the physical models, but it could be the radial coordinate $r$ in some of the models. Note that $Q = \{ q^i, i=1,\ldots, d \}$ is the configuration space from which it is possible to derive the corresponding tangent space $TQ= \{ q^i, {q'}^i \}$ on which the Lagrangian $\mathcal{L}$ is defined.

Taking the variation of $\mathcal{L}$ with respect to the generalized coordinates $q^i$, the Euler-Lagrange equations of motion become
\begin{equation}
\frac{d}{d \tau} \frac{\partial \mathcal{L}}{\partial {q'}^i} - \frac{\partial \mathcal{L}}{\partial q^i} = 0.  \label{e-l}
\end{equation}
The energy function associated with $\mathcal{L}$ is
\begin{equation}
E_{\mathcal{L}} = {q'}^k \frac{\partial \mathcal{L}}{\partial {q'}^k} - \mathcal{L}, \label{energy}
\end{equation}
which is also the Hamiltonian of the system. From a first order Lagrangian $\mathcal{L}= \mathcal{L} (r, q^k, {q'}^k)$ for stationary spacetimes, it follows the system of second-order ODEs of the form
\begin{equation}
{q''}^i = w^i (r, q^k, {q'}^k ). \label{ode}
\end{equation}

The Noether symmetry generator for the system of ODEs (\ref{ode}) is
\begin{eqnarray}
& & {\bf X} = \xi (r, q^k) \partial_r + \eta^i (r, q^k) \partial_{q^i} \, , \nonumber
\end{eqnarray}
if there exists a gauge function $h(r,q^k)$ and the Noether symmetry condition
\begin{equation}
{\bf X}^{[1]} \mathcal{L}+ \mathcal{L} ( D_r \xi )= D_r h \, , \label{ngs-cond}
\end{equation}
is satisfied. Here $D_r = \partial_r  + {q'}^k \partial_{q^k}$ is the total derivative operator and ${\bf X}^{[1]}$ is the first prolongation of Noether Symmetry generator ${\bf X}$, i.e.
\begin{eqnarray}
& & {\bf X}^{[1]} = {\bf X} + {\eta'}^k (r, q^{\ell}, {q'}^{\ell}) \partial_{ {q'}^k} \, , \label{first-pro}
\end{eqnarray}
where ${\eta'}^k (r, q^{\ell}, {q'}^{\ell}) = D_r \eta^k - {q'}^k D_r \xi$.
For every Noether symmetry generator, there is a \emph{conserved quantity} (a first integral) of the system of equations \eqref{ode} given by
\begin{equation}
I = - \xi E_{\mathcal{L}} + \eta^i \frac{\partial \mathcal{L}}{\partial {q'}^i} - h. \label{first-int}
\end{equation}

The most of the applications of Noether theorem to the extending theories of gravity are concerned with the following standard form of the Lagrangian
\begin{eqnarray}
& & \mathcal{L} = \mathcal{T} - V  \nonumber  \\ & & \quad = \frac{1}{2} \sigma_{i j} (r, q^k) {q'}^i {q'}^j + \gamma_j (r, q^k) {q'}^j -V(r, q^k), \label{lagr}
\end{eqnarray}
where $\mathcal{T}$ is the kinetic energy with a {\it kinetic metric} \cite{tp2011a}
\begin{equation}
ds_{\sigma}^2 = \sigma_{i j} dq^i dq^j \, ,
\end{equation}
for the configuration space, the indices $i,j,k,...$ run over the dimension of this space and $V(r, q^k)$ is the potential energy function. The function $\gamma_j (r, q^k)$ is a factor of the velocity term.

For the form of Lagrangian \eqref{lagr}, we obtain the first prolongation of the Noether symmetry generator ${\bf X}$ as
\begin{eqnarray} & & {\bf X}^{[1]} \mathcal{L} = - \xi V_{r} - \eta^k V_{,k} + \gamma_k \eta^k_{,r} + \left(  \pounds_{\bf \eta} \gamma_j - \xi_{,r} \gamma_j + \xi \gamma_{j,r} + \sigma_{ij} \eta^i_{,r}  \right) {q'}^j  \nonumber \\& & \qquad \qquad  + \frac{1}{2} \left( \pounds_{\bf \eta} \sigma_{ij} - 2 \xi_{,r} \sigma_{ij} + \xi \sigma_{ij,r} - 2 \xi_{,i} \gamma_j \right) {q'}^i {q'}^j  - \xi_{,k} \sigma_{ij} {q'}^i {q'}^j {q'}^k  \, ,  \label{1st-pro}
\end{eqnarray}
where $\pounds_{\bf \eta}$ is the Lie derivative operator along ${\bf \eta} = \eta^k \partial_{ q^k}$. Putting  (\ref{1st-pro}) into (\ref{ngs-cond}) together with $D_r \xi = \xi_{,r} + \xi_{,k} {q'}^k$ and $D_r h = h_{,r} + h_{,k} {q'}^k$, the Noether symmetry condition (\ref{ngs-cond}) becomes
\begin{eqnarray}
& & \xi_{,i} = 0, \quad \pounds_{\bf \eta} \gamma_i  + \xi \gamma_{i,r} + \sigma_{ij} \eta^j_{,r} - h_{,i} = 0, \label{ngs-cond-1} \\& & \qquad \pounds_{\bf \eta} \sigma_{ij} =(\xi_{,r}) \sigma_{ij} - \xi \sigma_{ij,r} \label{ngs-cond-2} \\ & & \eta^k V_{,k} + \left( V \xi \right)_{,r} - \gamma_i \eta^i_{,r} + h_{,r} = 0. \label{ngs-cond-3}
\end{eqnarray}
The above conditions explicitly yield the geometrical character of the Noether symmetry. Here,
$\xi_{,i} = 0$ implies $\xi = \xi(r)$.

\section{Noether symmetry in three-dimensional spacetime}
\label{sec:2}

The most general line element for the (2+1)-dimensional rotating black hole spacetime can be written in the form
\begin{equation}
ds^2 = - N(r)^2 dt^2 + \frac{dr^2}{A(r)^2} + P(r)^2 \left[ Q(r) dt + d\phi \right]^2  \, , \label{metric-1}
\end{equation}
where $N(r)$ is the lapse function and $Q(r)$ is the angular shift function. For this metric the Ricci scalar is given by
\begin{eqnarray}
& & R = - 2 A^2 \Big( \frac{N''}{N} + \frac{P''}{P}  +  \frac{A' N'}{A N}  + \frac{A' P'}{A P}  + \frac{N' P'}{N P} - \frac{P^2}{4 N^2} Q'^2 \Big), \label{ric-sc}
\end{eqnarray}
where the prime ($'$) represents the derivative with respect to $r$. Now, taking $f_R = \frac{d f}{d R}$ and $f_{R R} = \frac{d^2 f}{d R^2}$, one can obtain the following field equations
\begin{eqnarray}
& & f_R \left( \frac{N''}{N} + \frac{P''}{P} + \frac{A' N'}{A N} + \frac{A' P'}{A P}  - \frac{P^2 Q'^2}{2 N^2} \right) - f'_R \left( \frac{N'}{N} + \frac{P'}{P} \right) + \frac{f}{2 A^2} = 0 \, , \label{feq11}  \\ & & f_R \left( \frac{P''}{P} + \frac{A' P'}{A P} + \frac{N' P'}{N P} + \frac{P^2 Q'^2}{2 N^2} \right)  - f'_R \left( \frac{A'}{A} + \frac{N'}{N} \right) - f''_R + \frac{f}{2 A^2} = 0 \, ,  \label{feq22}  \\ & & f_R \left( \frac{ Q''}{Q} + \frac{3 P' Q'}{P Q} + \frac{A' Q'}{A Q}  - \frac{N' Q'}{N Q} \right) +  f'_R \frac{Q'}{Q} = 0  \, , \label{feq23} \\ & & f_R \left( \frac{ N''}{N} + \frac{A' N'}{A N} + \frac{N' P'}{N P} - \frac{P^2 Q'^2}{2 N^2} \right) -  f'_R \left( \frac{A'}{A} + \frac{P'}{P} \right) - f''_R + \frac{f}{2 A^2} = 0 \, ,   \label{feq33}
\end{eqnarray}
corresponding to $(rr), (\phi \phi), (t \phi)$ and $(tt)$ components of the Eq.\eqref{feqs}, respectively. After some calculations, it follows from the above field equations that
\begin{eqnarray}
& & f_R \frac{P''}{P} + f''_{R} + \left( \frac{A'}{A} - \frac{N'}{N} \right) \left( f'_R + f_R \frac{P'}{P} \right) = 0 , \qquad  \label{P-fR}
\end{eqnarray}
which will be useful to generate solutions for $f(R)$ theory of gravity.

In order for studying Noether symmetries we need to obtain a canonical Lagrangian of $f(R)$ gravity for the considered metric \eqref{metric-1}. Then, selecting the suitable Lagrange multiplier and integrating by parts, the Lagrangian $\mathcal{L}$ becomes canonical. Thus the point-like canonical Lagrangian of $f(R)$ gravity for $(2+1)$-dimensional black hole metric has the form
\begin{eqnarray}
& & \mathcal{L} = \frac{A P^3}{2 N} f_R Q'^2   + 2 f_R A N' P'  + 2 A P f_{RR} N' R'  + 2 A N f_{RR} P' R' + \frac{N P}{A} ( f -R f_R ). \,\, \qquad \label{lagr-fr}
\end{eqnarray}
The field equations obtained by variation of the action with respect to the metric coefficients $N, A, P, Q$ and $R$ take the form, respectively
\begin{eqnarray}
& & f_{R} \left( \frac{P''}{P} + \frac{A' P'}{A P} + \frac{P^2 Q'^2}{4 N^2} \right) + f'_{R} \left( \frac{A'}{A} + \frac{P'}{P} \right) + f''_R - \frac{1}{2 A^2} (f - R f_R) = 0, \label{feq-1}
\end{eqnarray}
\begin{eqnarray}
& & f_R \left( \frac{N' P'}{N P} + \frac{P^2 Q'^2}{4 N^2} \right) +  f'_R \left(  \frac{N'}{N} + \frac{P'}{P} \right)  - \frac{1}{2 A^2} (f - R f_R) = 0,  \label{feq-2}
\end{eqnarray}
\begin{eqnarray}
& & f_R \left( \frac{N''}{N} +  \frac{A' N'}{A N} - \frac{3 P^2 Q'^2}{4 N^2} \right) +  f'_R \left( \frac{A'}{A} + \frac{N'}{N} \right) + f''_R - \frac{1}{2 A^2} (f - R f_R) = 0, \label{feq-3}
\end{eqnarray}
\begin{eqnarray}
& & \left( \frac{ A P^3}{N} f_R Q' \right)' = 0, \label{feq-4}  \\& & f_{RR} \Big{[} 2 A^2 \Big( \frac{N''}{N} + \frac{P''}{P}  +  \frac{A' N'}{A N} + \frac{A' P'}{A P} + \frac{N' P'}{N P} - \frac{P^2}{4 N^2} Q'^2 \Big) +  R   \Big{]} = 0. \label{feq-5}
\end{eqnarray}
Note that the latter equation gives the Ricci scalar \eqref{ric-sc} if $f_{RR} \neq 0$. It is seen that by inserting the Ricci scalar $R$ given by \eqref{ric-sc} into the above equations \eqref{feq-1}-\eqref{feq-4}, one get the field equations \eqref{feq11}-\eqref{feq33}, which are the verification of variational field equations to be equivalent with the ones obtained from the tensorial form of the field equations \eqref{feqs}. Furthermore, the energy functional \eqref{energy} for the Lagrangian \eqref{lagr-fr}  becomes
\begin{eqnarray}
& & E_{\mathcal{L}} = 2 A N P \left[ f_R \left( \frac{N' P'}{N P} + \frac{P^2 Q'^2}{4 N^2} \right) +  f'_R \left( \frac{N'}{N} + \frac{P'}{P} \right) - \frac{1}{2 A^2} (f - R f_R) \right]. \label{energy-L}
\end{eqnarray}
After using the Ricci scalar (\ref{ric-sc}) in the above equation it yields $ E_{\mathcal{L}} = 0$ due to the field equation \eqref{feq11}.

Taking the configuration space variables as $q^i = (N, A, Q, P, R)$, $i= 1,2,3,4,5$, the kinetic metric for the Lagrangian (\ref{lagr-fr}) of $f(R)$ gravity is
\begin{eqnarray}
& & ds_{\sigma}^2 = \frac{A P^3}{2 N} f_R dQ^2 + 2 f_{R} A dN dP + 2 f_{RR} A P dN dR + 2 f_{RR} A N dP dR,  \label{km-fr}
\end{eqnarray}
and the potential is
\begin{equation}
V (N,A,P,R) = \frac{N P}{A} \left[ R f_R - f(R) \right].  \label{pot-fr}
\end{equation}
Then, the geometrical Noether symmetry conditions (\ref{ngs-cond-1})-(\ref{ngs-cond-3}) yield $\xi = \xi(r)$ and
\begin{eqnarray}
& &  f_{R}  \eta^3_{,A} =0, \quad h_{,A} =0, \quad 2 A \left( f_R \eta^4_{,r} + P f_{RR} \eta^5_{r} \right) - h_{,N} = 0, \nonumber \\& & A P^3 f_R \eta^3_{,r} - N h_{,Q} =0, \quad 2 A \left( f_R \eta^1_{,r} + N f_{RR} \eta^5_{,r} \right) - h_{,P} = 0,  \nonumber \\& & 2 A f_{RR} \left( P \eta^1_{,r} + N \eta^4_{,r} \right) - h_{,R}= 0, \nonumber\\& &  f_R \eta^4_{,N} + P f_{RR} \eta^5_{,N} = 0, \quad  f_{R} \eta^4_{,A} + P f_{RR}  \eta^5_{,A} = 0, \nonumber \\ & & 2 N \left(  f_R \eta^4_{,Q} + P f_{RR} \eta^5_{,Q} \right) + f_R P^3 \eta^3_{,N} = 0,  \nonumber \\ & &  f_R \eta^1_{,A} + f_{RR} N \eta^5_{,A} = 0, \quad f_{RR} \left( P \eta^1_{,A} + N \eta^4_{,A} \right) = 0, \nonumber \\
& & f_R \eta^1_{,P} + f_{RR} N \eta^5_{,P} = 0, \quad f_{RR} \left( P \eta^1_{,R} + N \eta^4_{,R} \right) = 0, \nonumber \\ & & f_R \left( P^3 \eta^3_{,P} + 2 N \eta^1_{,Q} \right) + 2 f_{RR} N^2 \eta^5_{,Q} = 0, \nonumber \\ & & f_R P^3 \eta^3_{,R} + 2 N f_{RR} \left( P \eta^1_{,Q} + N \eta^4_{,Q} \right) = 0, \label{ngs-eq}  \\& & f_R \left( - \frac{\eta^1}{N} + \frac{\eta^2}{A} + \frac{3 \eta^4}{P} + 2 \eta^3_{,Q} - \xi_{,r} \right) + f_{RR} \eta^5  = 0, \nonumber \\ & & f_R \left( \frac{\eta^2}{A} +  \eta^1_{,N} + \eta^4_{,P} - \xi_{,r} \right) + f_{RR} \left( \eta^5 + N \eta^5_{,N} + P \eta^5_{,P} \right) = 0, \nonumber \\
& & f_{RR} \left( \frac{\eta^1}{N} + \frac{\eta^2}{A}  + \frac{P}{N} \eta^1_{,P} + \eta^4_{,P} + \eta^5_{,R} - \xi_{,r} \right)  + \frac{f_R}{N} \eta^1_{,R} + f_{RRR} \eta^5 = 0, \nonumber\\ && f_{RR} \left( \frac{\eta^2}{A} + \frac{\eta^4}{P}  + \eta^1_{,N} + \frac{P}{N} \eta^4_{,N} + \eta^5_{,R} - \xi_{,r} \right)  + \frac{f_R}{P} \eta^4_{,R} + f_{RRR} \eta^5 = 0, \nonumber   \\& & V_{,N} \eta^1 + V_{,A} \eta^2 + V_{,P} \eta^4 + V_{,R} \eta^5 + V \xi_{,r} + h_{,r} = 0, \nonumber
\end{eqnarray}
where $V(N,A,P,R)$ is given by (\ref{pot-fr}), and the Noether symmetry generator to be find is
\begin{equation}
{\bf X} = \xi (r) \partial_r + \eta^1 \partial_N + \eta^2 \partial_A + \eta^3 \partial_Q + \eta^4 \partial_P + \eta^5 \partial_R \, .
\end{equation}
For this generator, the corresponding first integral \eqref{first-int} of the $f(R)$ Lagrangian \eqref{lagr-fr} is
\begin{eqnarray}
& & I = -\xi E_{\mathcal{L}} + 2 A \left( f_R P \right)' \eta^1 +  \frac{A P^3}{N} f_R Q' \eta^3  + 2 A \left( f_R N \right)' \eta^4 + 2 A f_{RR} \left( N P \right)' \eta^5.
\end{eqnarray}
The function $h$ of equation \eqref{ngs-eq} is assumed constant, unless otherwise stated.

It follows from the Noether symmetry equations given by \eqref{ngs-eq} that for arbitrary form of the function $f(R)$, there are {\it four} Noether symmetries
\begin{eqnarray}
& & {\bf X}_1 = \partial_Q, \quad {\bf X}_2 = N \partial_N +  2 Q \partial_Q - P \partial_P, \label{X12} \\& &  {\bf X}_3 = N Q \partial_N +  \left( Q^2 + \frac{N^2}{P^2} \right) \partial_Q - P Q \partial_P,  {\bf X}_4 =  \partial_r, \qquad  \label{X34}
\end{eqnarray}
with the Lie algebra
\begin{equation}
\left[{\bf X}_1,{\bf X}_2 \right] = 2 {\bf X}_1, \, \left[{\bf X}_1,{\bf X}_3 \right] = {\bf X}_2, \left[{\bf X}_2,{\bf X}_3 \right] = 2 {\bf X}_3  . \label{alg-1}
\end{equation}
Then the corresponding first integrals of ${\bf X}_1, {\bf X}_2, {\bf X}_3$ and ${\bf X}_4$ are
\begin{eqnarray}
& & I_1 = \frac{A P^3}{N} f_R Q' , \quad  I_2 = 2 I_1 Q + 2 A N^2 f_R  \left( \frac{P}{N} \right)' \, , \quad  \label{fint-1}  \\ & &  I_3 = I_1 \left( \frac{N^2}{P^2} - Q^2 \right) + I_2 Q,  \quad I_4 = - E_{\mathcal{L}},  \label{fint-2}
\end{eqnarray}
where $I_1, I_2, I_3, I_4$ are constants of motion, and $I_4$ vanishes due to the Hamiltonian constraint $E_{\mathcal{L}} = 0$.

It is possible to find solutions to the Noether symmetry equations \eqref{ngs-eq} where the form of $f$ is $f(R) = f_{0} R^{n}$ with $f_0$ and $n$ being constants. In this case, there are {\it five} Noether symmetries ${\bf X}_1, {\bf X}_2, {\bf X}_3, {\bf X}_4,$ given above by \eqref{X12} and \eqref{X34}, and additionally
\begin{equation}
{\bf X}_5 = N \partial_N - \frac{2 A}{2 n-1} \partial_A + P \partial_P - \frac{4 R}{2n-1} \partial_R  \, , \label{X5-r2}
\end{equation}
which gives rise to the first integral
\begin{eqnarray}
& & I_5 = 2 A \left[ \frac{(3-2 n)}{(2 n-1)} f_R \left( N P \right)' +  2 N P f'_R \right] ,  \label{fint1}
\end{eqnarray}
where $f_R = f_0 n R^{n-1}$, $f'_R = f_0 n (n-1) R^{n-2} R'$ and $n \neq 1/2$. For $n=1/2$, the vector field ${\bf X}_5$ is found from \eqref{ngs-eq} such that
\begin{equation}
{\bf X}_5 = A \partial_A + 2 R  \partial_R  \, , \label{X5-r2-2}
\end{equation}
with the first integral
\begin{eqnarray}
& & I_5 = - \frac{f_0 A}{\sqrt{R}} \left( N P \right)'  \, .  \label{fint1-2}
\end{eqnarray}
Here the Lie algebra of these five Noether symmetries are the same as \eqref{alg-1}.

One also finds that there are {\it seven} Noether symmetries satisfying the Noether symmetry equations \eqref{ngs-eq}, which are ${\bf X}_1, {\bf X}_2, {\bf X}_3, {\bf X}_4$ given in \eqref{X12} and \eqref{X34}, and
\begin{eqnarray}
& & {\bf X}_5 = \frac{1}{N} \partial_N + \frac{A}{N^2} \partial_A \, , \label{X5-r3-1} \\
& & {\bf X}_6 = \frac{Q}{N} \partial_N + \frac{A Q}{N^2} \partial_A + \frac{1}{P^2} \partial_Q \, , \label{X5-r3-2} \\ & & {\bf X}_7 = \frac{Q^2}{2 N} \partial_N + \frac{A ( P^2 Q^2 - N^2)}{2 N^2 P^2} \partial_A + \frac{Q}{P^2} \partial_Q - \frac{1}{2 P} \partial_P \, , \label{X5-r3-3}
\end{eqnarray}
while the form of $f$ becomes $f(R) = R - 2 \Lambda $, where $\Lambda$ is a constant. Thus the first integrals for ${\bf X}_5, {\bf X}_6, {\bf X}_7 $ are
\begin{eqnarray}
& & I_5 = \frac{2 A P'}{N}, \quad  I_6 = \frac{I_1}{P^2} + I_5 Q , \label{fint2-1} \\& &  I_7 = I_6 Q - I_5 \frac{Q^2}{2} - \frac{A}{P} N' . \label{fint2-2}
\end{eqnarray}
Furthermore, for this case the Hamiltonian constraint $E_{\mathcal{L}} = 0$ yields
\begin{equation}
A = - \frac{1}{I_3 N'} \left( \frac{I_1^2}{2 P^3} + 2 \Lambda P \right),
\end{equation}
with $I_3 N' \neq 0$. Also, the corresponding Lie algebra of Noether symmetries ${\bf X}_1, ..., {\bf X}_7$ has the following non-vanishing commutators
\begin{eqnarray}
& & \left[{\bf X}_1,{\bf X}_2 \right] = 2 {\bf X}_1, \, \left[{\bf X}_1,{\bf X}_3 \right] = {\bf X}_2, \left[{\bf X}_1,{\bf X}_6 \right] = {\bf X}_5, \nonumber \\ & & \left[{\bf X}_1,{\bf X}_7 \right] = {\bf X}_6, \left[{\bf X}_2,{\bf X}_3 \right] = 2 {\bf X}_3, \left[{\bf X}_2,{\bf X}_5 \right] = -2 {\bf X}_5, \qquad \label{alg-2} \\ & & \left[{\bf X}_2,{\bf X}_7 \right] = 2 {\bf X}_7,  \left[{\bf X}_3,{\bf X}_5 \right] = -2 {\bf X}_6,  \left[{\bf X}_3,{\bf X}_6 \right] = - 2 {\bf X}_7 \, . \nonumber
\end{eqnarray}

In the following section, we will use the above Noether symmetries to derive exact solutions for the corresponding (2+1)-dimensional $f(R)$ theories of gravity.

\section{Exact solutions}
\label{sec:3}

For any form of $f(R)$, we consider the first integrals $I_1, \, I_2 , \, I_3$ given in \eqref{fint-1} and \eqref{fint-2}. For the sake of simplicity, we take the constants of motion as $I_1 = a,\,\, I_2 = b$ and $I_3 = c$. It follows from those of the first integrals that
\begin{eqnarray}
& & N^2 = P^2 \left( \frac{c - b Q}{a} + Q^2 \right), \label{sl-1}  \\ & & Q' = \frac{ a N}{f_R A P^3},  \label{sl-2} \\ & &  \left( \frac{P}{N} \right)' = \frac{ b - 2 a Q}{2 f_R A N^2}  , \label{sl-3}
\end{eqnarray}
where $f_R \neq 0$. Let us now consider the cases having five and seven Noether symmetries obtained before.

\bigskip

{\bf Case (i):} In this case, the form of $f$ is $f(R) = R - 2 \Lambda$ and there exist seven Noether vectors given in \eqref{X12}, \eqref{X34}, \eqref{X5-r3-1}, \eqref{X5-r3-2} and \eqref{X5-r3-3}. Here, $\Lambda$ is obviously identified as the cosmological constant. Then, taking $I_5 = k$ and rearranging the first integrals for these Noether symmetries, we obtain
\begin{eqnarray}
& & N^2 = \frac{ a Q^2 - b Q + c}{I_6 - k Q}, \quad P^2 = \frac{a}{I_6 - k Q}, \label{i-1}
\\ & & P' = \frac{k N}{2 A}, \quad Q' = \frac{a N}{A P^3}, \label{i-2} \\ & & 2 P A N' - k N^2 - 2 a Q + b = 0,  \label{i-3} \\ & & A N' + P \left( \frac{k}{2} Q^2 - I_6 Q + I_7 \right) = 0, \label{i-4} \\ & & 2 k A N' + \frac{a^2}{P^3} + 4 \Lambda P = 0, \label{i-5}
\end{eqnarray}
with $Q \neq I_6 / k$. In order to solve the above system of differential equations one can define a function $F(r)$
\begin{equation}
F(r) = \int{ \frac{N}{A} dr} , \label{def-F}
\end{equation}
which means that $N(r) = A(r) F'(r)$. Then, the metric functions $P$ and $Q$ can be found via \eqref{i-2} as
\begin{eqnarray}
P = \frac{k}{2} F(r) + d_1, \quad Q = - \frac{4 a}{k ( k F(r) + 2 d_1)^2} + d_2, & & \,\,  \label{i-PQ}
\end{eqnarray}
where $d_1, d_2$ are integration constants. Thus, the first relation in Eq. \eqref{i-1} yields
\begin{equation}
N^2 = \frac{a^2}{ k^2 \left( \frac{k}{2} F(r) + d_1 \right)^2} + d_3 \left( \frac{k}{2} F(r) +  d_1 \right)^2 + N_0 \, , \label{i-N2}
\end{equation}
where $d_3 = ( c - b d_2 )/a  +  d_2^2$, $N_0 \equiv ( b - 2 a d_2)/k$ and $k \neq 0$.  The remaining equations \eqref{i-3}-\eqref{i-5} give the Noether constants as follows
\begin{eqnarray}
& & I_6 = k d_2, \qquad I_7 = \frac{k}{2} \left( d_2^2 - d_3\right),  \label{c-i-1} \\ & & d_3 = - \frac{4 \Lambda }{k^2}, \qquad c = b d_2 + a ( d_3 - d_2^2) . \label{c-i-2}
\end{eqnarray}
It has to be noted that the solutions \eqref{i-PQ} and \eqref{i-N2} of the metric functions $N, P$ and $Q$ are found in terms of an arbitrary function $F(r)$ under the condition $N(r) = A(r) F'(r)$. Therefore, if one choose the function $F(r)$, then one can find the metric functions explicitly. The Eq.\eqref{P-fR} for this case becomes
\begin{equation}
\frac{P''}{P} + \left( \frac{A'}{A} - \frac{N'}{N} \right) \frac{P'}{P} = 0 , \label{P-fR-i}
\end{equation}
which is easily satisfied by taking $P = \frac{k}{2} F(r) + d_1$ and $N(r) = A(r) F'(r)$ for any form of $F(r)$ with the condition $F'(r) \neq 0$.
In the following we conclude this case by pointing out some examples.

Firstly, we simply take $A = N$ and $F (r) = 2 (r -d_1)/k$, which gives rise to the well-known form of BTZ metric \cite{btz}. Then, it follows from \eqref{i-PQ} and \eqref{i-N2} that one can explicitly write
\begin{equation}
P(r) = r, \quad Q(r) = - \frac{a}{ k r^2} + d_2, \label{i-2-PQ}
\end{equation}
and
\begin{equation}
N^2 = \frac{ a^2}{ k^2 r^2} + d_3\, r^2  + N_0 \, . \label{i-2-N2}
\end{equation}
Here, the Noether constants are the same ones as in \eqref{c-i-1} and \eqref{c-i-2}.
Taking the special values of parameters such as $d_2 = 0, N_0 = -M, a = k J / 2$ and $d_3 = \ell^{-2}$ in the obtained metric functions $N, P$ and $Q$ one can reach the BTZ black hole solution \cite{btz}
\begin{equation}
N(r)^2 = - M + \frac{r^2}{\ell^2} + \frac{J^2}{4 r^2} \, , \quad P(r)= r \, , \quad Q(r) = -\frac{J}{2 r^2}, \label{btz-orj}
\end{equation}
where $M = -b/k$ and $J = 2 a /k$ are the mass and angular momentum of the black hole, respectively. Here, the other Noether constants become $I_6 = 0, I_7 = - k \ell^{-2} /2,  d_3 = - 4 \Lambda /k^2 $ and $c = k J \ell^{-2}/2$. It should be pointed out that not only the mass and angular momentum of black hole but also the cosmological constant are Noether constants, i.e. $b = - k M, a = k J /2$ and $I_7 = 2 \Lambda /k$.

Secondly, now let us consider the special case where the solution is asymptotically Lifshitz black hole \cite{L1,L2}. For this aim, we use the ansatz $N = \left( \frac{r}{\ell} \right)^{z} H(r)$ and $A = \frac{r}{\ell} H(r)$ which yields $F(r)= \frac{\ell}{z} \left( \frac{r}{\ell} \right)^{z}$ from the relation \eqref{def-F}, where $H(r)$ is a function of the radial coordinate, and $z$ is the dynamical critical exponent. The $z=1$ value corresponds to the standard scaling behavior of conformal invariant solutions. Thus, one can write the metric functions
\begin{equation}
P = \frac{k \ell}{2 z} \left( \frac{r}{\ell} \right)^{z} + d_1 \, , \quad  Q = - \frac{4 a z^2}{ k \left[ k \ell \left( \frac{r}{\ell} \right)^{z} + 2 d_1 z \right]^2} + d_2, \label{i-3-PQ}
\end{equation}
and
\begin{equation}
N^2 = \frac{4 a^2 }{ k^2 \left[ \frac{k \ell}{z}  \left( \frac{r}{\ell} \right)^{z} + 2 d_1 \right]^2} + \frac{d_3}{4}  \left[ \frac{k \ell}{z} \left( \frac{r}{\ell} \right)^{z} + 2 d_1  \right]^2  + N_0 \, , \label{i-3-N2}
\end{equation}
that gives
\begin{eqnarray}
& & H^2 = \frac{4 a^2 }{ k^2 \left( \frac{r}{\ell} \right)^{2 z} \left[ \frac{k \ell}{z}  \left( \frac{r}{\ell} \right)^{z} + 2 d_1 \right]^2} + \frac{d_3}{4}  \left[ \frac{k \ell}{z}  + 2 d_1 \left( \frac{r}{\ell} \right)^{-z} \right]^2  + N_0 \left( \frac{r}{\ell} \right)^{-2 z} \, . \label{i-3-H2}
\end{eqnarray}
It is easily found that for $z > 1$ the function $H(r)$ given by \eqref{i-3-H2} obey $ \lim\limits_{r \to \infty} H(r) \rightarrow 1$  when $\Lambda = - z^2 / \ell^2$.

\bigskip

{\bf Case (ii):} For this case, we consider the \emph{five} Noether symmetries given by \eqref{X12}, \eqref{X34} and \eqref{X5-r2} where the function $f(R)$ has the form $f(R) = f_0 R^n$ due to the Noether symmetry equations. The corresponding conserved quantities (first integrals of motion) yield \eqref{sl-1} and
\begin{eqnarray}
& & \frac{N' P'}{N P} + \frac{P^2 Q'^2}{4 N^2}  + (n-1)\frac{R'}{R} \left( \frac{N'}{N} + \frac{P'}{P} \right) + \frac{(n-1) R}{2 n A^2} = 0 \, , \label{ii-eq1} \\ & & Q' = \frac{ a N R^{1-n}}{f_0 n A P^3}, \label{ii-eq2}  \\ & & N P' - P N' + \frac{ (2 a Q - b) R^{1-n}}{2 f_0 n A}= 0 \, ,  \label{ii-eq3}  \\ & & \frac{(3-2n)}{(2n-1)} \left( N P \right)' +  2 (n-1) N P \frac{R'}{R} = \frac{ k R^{1-n}}{2 f_0 n A} , \label{ii-eq4}
\end{eqnarray}
where $n\neq 1/2$. Then, we will search exact solution of the above first integral equations to find the metric coefficients $A, N, P$ and $Q$.

First of all, we assume $A = N$ to arrive at solutions form the above differential equations \eqref{ii-eq1}-\eqref{ii-eq4}. Then, the Eq. \eqref{P-fR} becomes
\begin{equation}
\frac{P''}{P} + (n-1) \left[ \frac{R''}{R} + (n-2) \frac{R'^2}{R^2} \right] = 0. \label{P-fR-ii}
\end{equation}
This equation relates the metric function $P$ and the Ricci scalar $R$, and if one chooses $P$, one can find $R$ by solving \eqref{P-fR-ii}, or vice versa. For example, for $P(r) = P_1 r^{(1+ \alpha)/2} + P_2 r^{(1 - \alpha)/2}$,  the Eq. \eqref{P-fR-ii} has a solution of the Ricci scalar $R$ as
\begin{equation}
R(r)=  \left[ R_1 r^{(1+ \beta)/2} + R_2 r^{(1 - \beta)/2}  \right]^{\frac{1}{n-1}} \, , \, n \neq 1, \quad
\end{equation}
where $R_1$ and $R_2$ are integration constants, $\alpha$ and $\beta$ are real constant parameters having the property $\alpha^2 + \beta^2 = 2$. It is very difficult to solve the Eqs.\eqref{ii-eq1}-\eqref{ii-eq4} using these form of $P$ and $R$. So we need to simplify the form of $P$ and $R$ choosing the coefficients $P_1, P_2, R_1$ and $R_2$. Now, taking $P_1 = 1$ and $P_2 = R_2 = 0$, that is, $P(r)= r^{(1+ \alpha)/2}$ and $R(r) =  K_1  r^{\frac{(1 + \beta)}{2(n-1)}} $ with $K_1 = R_1^{1/(n-1)}$, we find from the Eqs. \eqref{ii-eq1}-\eqref{ii-eq4} that
\begin{eqnarray}
& & Q(r) = - \frac{ 2 a \, r^{-(3 \alpha + \beta + 2)/2}}{ f_0 nR_1 (3 \alpha + \beta + 2)} + q_1, \label{ii-sl11}  \\ & & A(r) = \frac{ 2 a}{ f_0 n R_1 ( 3 \alpha + \beta + 2)} r^{ -( 2 \alpha + \beta +1)/2 }, \label{ii-sl12}
\end{eqnarray}
where the conditions $k=0, b = 2 a q_1$ and $c = a q_1^2$ have to be satisfied, and
\begin{eqnarray}
& & a^2 =  \frac{2 f_0^2  n(n-1)( 3 \alpha + \beta +2)^2}{ ( \alpha + \beta)( 3 \beta -\alpha + 4)}  R_1^{ \frac{2n-1}{n-1}},  \label{ii-K1}
\\ & & \alpha = -\frac{ ( 28 n^2 - 52 n + 23) }{ 20 n^2 - 36 n + 17} \,  , \quad
\beta = -\frac{( 4 n^2 + 4 n - 7)}{20n^2 - 36n + 17}. \qquad \label{ii-a-b}
\end{eqnarray}
Here, using \eqref{ii-a-b} in the term $3 \alpha + \beta + 2$ appeared at \eqref{ii-sl11} and \eqref{ii-sl12}, it gives rise to the constraint $n \neq \frac{1}{2}, \frac{7}{6}$. For $n=5/6$, we observe that $k$ needs not to be vanish, and one gets $\alpha = \beta = 1$ from \eqref{ii-a-b}. Then, the metric functions $P, Q$ and $A=N$ given as
\begin{eqnarray}
& & P(r) = r, \quad Q(r) = - \frac{a}{3 R_1 r^3} + q_1,  \label{ii-sl21}  \\ & &
A(r)^2 = \frac{a^2}{9 R_1^2 \, r^4} + \frac{ (b - 2 a q_1)}{3 R_1 \, r} \, , \label{ii-sl22}
\end{eqnarray}
solve the Eqs. \eqref{ii-eq1}-\eqref{ii-eq4} under the conditions $c = b q_1 - a q_1^2$ and $k = 2 b - 4 a q_1$. Here the Ricci scalar is $R(r)= K_1/ r^6$ with $K_1 = -5 a^2 / (6 R_1^2)$. Furthermore, for $n=3/2$,  one can get the following solution of Eqs. \eqref{ii-eq1}-\eqref{ii-eq4} taking $\alpha = 1, \beta=-1$ and $R_2 = 0$,
\begin{equation}
P(r) = r, \quad Q(r) = - \frac{a}{2 R_1 r^2} + q_1,  \label{ii-sl31}
\end{equation}
\begin{equation}
A(r)^2 =  \frac{(b - 2 a q_1)}{2 R_1} + \frac{a^2}{4 R_1^2 \, r^2} +  \left( a q_1^2  - b q_1 + c \right) \frac{r^2 }{a}  \, , \label{ii-sl32}
\end{equation}
with $k=0$ and $R(r) = 6 ( b q_1 - a q_1^2 - c)/a $ = const. If one assumes $q_1 = 0, a = R_1 J, b = - 2 R_1 M $ and $c  = a/\ell^2$ in the above metric functions, one can arrive the original BTZ black hole solution. This interestingly means that $f(R)= f_0 R^{3/2}$ theory of gravity gives also rise to the BTZ black hole solution. For $n \neq 3/2$, one can find the solution when $\alpha = \beta = -1$ and $R_2 = 0$ such as
\begin{eqnarray}
& &  P(r) = r , \quad  Q(r)= - \frac{a}{ 2 R_1 \, r^2 } + q_1, \qquad \label{ii-sl41} \\ & & A(r)^2 =  \frac{k (1 - 2n)}{2 R_1 ( 2n -3)} + \frac{a^2}{ 4 R_1^2 \, r^2}  \,  , \label{ii-sl42}
\end{eqnarray}
through the conditions $b = 2 a q_1 + k (1 - 2n)/(2n -3)$ and $c = b q_1 - a q_1^2$, which yields that the Ricci scalar vanishes, i.e. $R = 0$. Therefore, we conclude that the metric for $n \neq 3/2$, which includes the metric for $n=7/6$, is a vacuum solution. Furthermore, for $n=1/2$, we have to use the first integral \eqref{fint1-2} instead of \eqref{fint1}, or \eqref{ii-eq4}. Then, one can obtain a vacuum solution, namely $R = 0$, in which the metric functions have the form
\begin{eqnarray}
& &  P(r) = r , \quad  Q(r)= - \frac{a}{ 2 \, r^2 } + q_1 \, , \label{ii-sl51} \\ & & A(r) = N(r) = \frac{a}{ 2 \, r } \, , \label{ii-sl52}
\end{eqnarray}
where $k = 0, b = 2 a q_1$ and $c = a q_1^2 $. Therefore, one can conclude that the metric for $n=1/2$ is also a vacuum solution.

Now, we look for the solutions of \eqref{ii-eq1}-\eqref{ii-eq4} under the assumption of $A \neq N$, where the Eq. \eqref{P-fR} reads
\begin{eqnarray}
& & \frac{P''}{P} + (n-1) \left[ \frac{R''}{R} + (n-2) \frac{R'^2}{R^2} \right] + \left( \frac{A'}{A} - \frac{N'}{N} \right) \left[ \frac{P'}{P} + (n-1) \frac{R'}{R} \right] = 0.  \label{P-fR-ii-2}
\end{eqnarray}
One can explicitly solve Eq.\eqref{ii-eq4} for $k=0$, finding that
\begin{equation}
N = \frac{N_0}{P} R^{ \frac{2(n-1)(2n-1)}{2n-3} }, \label{N-ii-2}
\end{equation}
where $N_0$ is an integration constant, and $n \neq 3/2$. Using \eqref{N-ii-2} in \eqref{sl-1}, the metric function $Q$ is obtained as
\begin{equation}
Q = \frac{b}{2 a} \pm \frac{N_0}{P^2} R^{ \frac{2(n-1)(2n-1)}{2n-3} }, \label{Q-ii-2}
\end{equation}
for $c = b^2 / 4 a$. Then, assuming $P(r) = r^{ ( \alpha +1)/2}$, the Eqs. \eqref{ii-eq1}-\eqref{ii-eq3} and \eqref{P-fR-ii-2} have the following solutions
\begin{eqnarray}
& & R(r) = \left( K_2 r \right)^{ -\frac{2(\alpha+1)}{ 2n-1} },  \quad  A(r)= A_0 r^{ \frac{ 2(n-1) -\alpha}{2n-1}}, \label{R-A-case-ii-6}
\end{eqnarray}
where $A_0$ is
\begin{equation}
A_0 = -\frac{ a (2n-3) K_2^{ \frac{2(\alpha+1)(n-1)}{2n-1}} }{ f_0 n (\alpha+ 1) (6n-7) }, \label{A0-case-ii-6}
\end{equation}
and
\begin{equation}
a = \frac{ f_0}{K_2^{\alpha +1}} \sqrt{ \frac{ n(2n-1)(6n-7)}{ 8 (1-n)} }  \label{a-case-ii-6}
\end{equation}
with $n \neq 0,1,1/2, 7/6$. Putting the latter form of $P$ and $R$ into $N$ and $Q$ given in \eqref{N-ii-2} and \eqref{Q-ii-2}, then $N$ and $Q$  take the form
\begin{equation}
N(r) = \frac{N_0 \, r^{ -\frac{(\alpha +1)(10n-11)}{2(2n-3)}} }{ K_2^{ \frac{ 4(\alpha+1) (n-1)}{2n-3} } } , \label{N2-ii-2}
\end{equation}
and
\begin{equation}
Q (r) = \frac{b}{2 a} \pm \frac{N_0 \, r^{ \frac{-(\alpha +1)(6n-7)}{2n-3} }  }{ K_2^{ \frac{ 4(\alpha+1) (n-1)}{2n-3} } } \, . \label{Q2-ii-2}
\end{equation}
For $n=3/2$, if $k=0$ and $R = R_0 $ = const., we can find the following exact solution of metric functions
\begin{eqnarray}
& & P(r)= r^{\nu}, \quad Q(r) = - \frac{ a }{ q_0 r^{2 \nu} } + q_1, \label{P-case-ii-7}   \\ & & A(r)^2 = \frac{r^2}{\nu^2} \left[  \frac{ a q_1^2 - b q_1 + c}{ a} + \frac{a^2}{ q_0^2 r^{4 \nu}} + \frac{ b- 2 aq_1}{ q_0 \, r^{2 \nu}} \right], \qquad \label{A-case-ii-7}  \\ & & N(r)^2 = \nu^2 r^{ 2 (\nu -1)} A(r)^2, \label{N-case-ii-7}
\end{eqnarray}
in which $q_0 = 3 f_0 \sqrt{R_0}$, $R_0 = 6 ( b q_1 - a q_1^2 - c)/a$ and $\nu$ is a constant parameter. We note that for $\nu = 1$ the latter exact solution reduces to \eqref{ii-sl31} and \eqref{ii-sl32} where $A = N$.

\bigskip

{\bf Case (iii):} In this case, we take the \emph{four} Noether symmetries given in \eqref{X12} and \eqref{X34}, which imposes an arbitrary form of the function $f(R)$. The first integrals of this case are given by \eqref{sl-1}, \eqref{sl-2}, \eqref{sl-3} and $E_{\mathcal{L}} = 0$ which gives
\begin{eqnarray}
& & f_R \left( \frac{N' P'}{N P} + \frac{P^2 Q'^2}{4 N^2} \right) +  f'_R \left( \frac{N'}{N} + \frac{P'}{P} \right) - \frac{1}{2 A^2} (f - R f_R) = 0. \label{fint-iii}
\end{eqnarray}

For this case, we can firstly choose that $P(r) = r$ and $A=N$. Then, the Eq. \eqref{P-fR} becomes $f''_R = 0$ which has a solution $f_R = R_1 r + R_2$, where $R_1, R_2$ are integration constants. This information is sufficient to solve the Eq. \eqref{sl-2} to find $Q$ as follows
\begin{equation}
Q(r) = \frac{ a R_1^2}{R_2^3} \ln \left( \frac{r}{R_1 r + R_2} \right) + \frac{a}{R_2} \left( \frac{R_1}{R_2 r} - \frac{1}{2 r^2} \right) + Q_1,  \label{Q-r}
\end{equation}
where $Q_1$ is an integration constant. Furthermore, the field Eq. (\ref{feq-5}) yields
\begin{equation}
(A^2)'' + \frac{2}{r} (A^2)' - \frac{a^2}{2 r^4 (R_1 r + R_2)^2} + R = 0,  \label{feq-444}
\end{equation}
for $f_{R R} \neq 0$, i.e. $R_1 \neq 0$ in this case. Thus, one need to select the function $R(r)$ to find a solution of the above ordinary differential equation, i.e. the function $A(r)^2$.
Due to the differential equation $f_R = R_1 r + R_2$, using the ansatz $R(r) = K_0 r^m$ proposed in \cite{darabi2013}, the $f(R)$ takes the form
\begin{equation}
f(R) = \frac{m R_1 }{(m+1) K_0^{1/m} }  R^{\frac{m+1}{m}} + R_2 R + R_0,  \label{fR1}
\end{equation}
where $K_0$ is a dimensional constant, $R_0$ is a constant of integration, and $m \neq -1$.
Putting $R(r) = K_0 r^m$ into Eq. \eqref{feq-444}, we can integrate \eqref{feq-444} to find
\begin{eqnarray}
& & A^2 (r) = \frac{ a^2 R_1}{R_2^3} \left( \frac{1}{r} + \frac{3 R_1}{2 R_2} \right) \ln \left( \frac{ r }{R_1 r + R_2} \right) + \frac{a^2}{4 R_2^2 r^2}  + \left( \frac{R_1 a^2 }{2 R_2^3}  + A_1 \right)\frac{1}{r}   - \frac{K_0 r^{m+2}}{(m+2)(m+3)} + A_2, \label{N2-r}
\end{eqnarray}
where $A_1, A_2$ are constants of integration, and $m \neq -2,-3$. Note that the above solution \eqref{N2-r} was reported before \cite{darabi2013}. Unfortunately, the metric functions $P(r) = r$, $Q(r)$ by \eqref{Q-r} and $N(r)$ by \eqref{N2-r} are not solve the first integral relations \eqref{sl-3} and \eqref{fint-iii}. Thus, those of metric functions are not solutions of the field equations \eqref{feq-1}-\eqref{feq-5} unless one takes that $R_1 =0, A_1 = 0, m= 0, R_2=1, K_0 = -3 R_0, a= J, A_2 = -M$, and $R_0 = 2 \ell^{-2}$ which is just the BTZ black hole solution. Let us discuss why this happened.
The point-like Lagrangian used in the Ref. \cite{darabi2013} is poorly described one because of that it gives rise to incomplete variational field equations of motion, in which there are \emph{three} variational equations of motion, the Eqs. (11), (12) and (12) of \cite{darabi2013}. We observe that the reason for missing some of field equations after the variation of the Lagrangian given in the Ref. \cite{darabi2013} is the lack of taking variation with respect to some metric functions that depend explicitly on $r$. To tackle this problem, we have written all of the metric coefficients in terms of implicit functions of $r$ which are $N, A, P$ and $Q$. Afterwards, in order to recover the underlying field equations, we have varied the obtained Lagrangian \eqref{lagr-fr} with respect to these implicit functions, i.e. the metric coefficients, and the Ricci scalar $R$ that implicitly depend on $r$. So, we have arrived the fact that the number of variational equations of motion is \emph{five} given by \eqref{feq-1}-\eqref{feq-5}, \emph{not three} as in the Ref. \cite{darabi2013} .

Now, we want to obtain exact solutions for $N \neq A$. For this purpose, by employing $P(r)=r$ and $N(r) = ( r/r_0 )^z A(r)$ which gives a Lifshitz-like 3-dimensional spacetime, the Eq. \eqref{P-fR} has the following solution
\begin{equation}
f_R = D_1 r^{\frac{1}{2} \left( z + 1 + \alpha\right)} +  D_2 r^{\frac{1}{2} \left( z + 1 - \alpha \right)}, \label{fR1-iii}
\end{equation}
where $D_1, D_2$ are integration constants, $z$ is a real number so called Lifshitz-like parameter, $r_0$ is an arbitrary (positive) length scale, and $\alpha \equiv \sqrt{ z^2 + 6 z +1}$ which yields that $z < z_1 = -3 - 2 \sqrt{2}$ and $z > z_2 = -3 + 2 \sqrt{2}$. It is very difficult to find out a general form of $f(R)$ from \eqref{fR1-iii} for arbitrary $z$. Therefore, we will take $z=-6$ as an example. Then, the Eq. \eqref{fR1-iii} becomes
\begin{equation}
f_R = \frac{D_1}{r^2} +  \frac{D_2}{r^3},  \label{fR2-iii}
\end{equation}
which can be written in the form
\begin{equation}
f' = \left( \frac{D_1}{r^2} +  \frac{D_2}{r^3} \right) R',  \label{fR2-2-iii}
\end{equation}
where $f' = f_R R'$. If $D_1 \neq 0$ and $D_2 = 0$, then the solution of Eq.\eqref{sl-2} is
\begin{equation}
Q(r) = -\frac{a}{6 D_1} \left( \frac{r_{_0}}{r} \right)^6 +  Q_1 \, ,  \label{Q1-iii}
\end{equation}
where $Q_1$ is an integration constant. Using the latter $Q$, $P(r)= r$ and $N(r) = \left( r_0 / r \right)^6 A(r)$ in the Eqs.\eqref{sl-1} and \eqref{sl-3}, one can obtain the lapse function $A(r)^2$ as
\begin{eqnarray}
& & A(r)^2 = r^2 \left[ -A_1 r^{12} +  \frac{\left( b - 2 a Q_1 \right)}{6 D_1 r_0^6}  r^6  + \frac{a^2}{ 36 D_1^2} \right] \, , \qquad  \, \label{A1-iii}
\end{eqnarray}
with the constraint $c = b Q_1 - a ( A_1 r_0^{12} + Q_1^2)$, where $A_1$ is a constant of integration. Putting the metric functions obtained here to the definition given in  \eqref{ric-sc}, we have the Ricci scalar
\begin{equation}
R(r) = 30 A_1 r^{12} -\frac{2 a^2}{3 D_1^2}   \, ,  \label{R1-iii}
\end{equation}
which gives $f(r) = 36 A_1 D_1 r^{10}$ from \eqref{fR2-2-iii}. Thus,  one can easily find $f(R)$ as $f(R) = f_0 \left( R + R_0 \right)^{5/6}$, where $f_0 = 36 D_1 \left( A_1^{1/5} /30 \right)^{5/6}$ and $R_0 = 2 a^2 / 3 D_1^2 $. Note that the Eq.\eqref{fint-iii} and the field equations are equivalently satisfied for this solution. If $D_1 = 0$ and $D_2 \neq 0$, then using  $P(r)= r$ and $N(r) = \left( r_0 / r \right)^6 A(r)$ in the first integrals \eqref{sl-1}, \eqref{sl-2} and \eqref{sl-3}, we have the following solutions
\begin{equation}
Q(r) = -\frac{a r_0^6}{5 D_2 r^5} +  Q_2  \, ,  \label{Q2-iii}
\end{equation}
\begin{eqnarray}
& & A(r)^2 = r^4 \left[ -A_2 r^{10} +  \frac{\left( b - 2 a Q_2 \right)}{5 D_2 r_0^6}  r^5  + \frac{a^2}{ 25 D_2^2} \right] \, , \qquad  \, \label{A2-iii}
\end{eqnarray}
with the constraint $c = b Q_2 - a ( A_2 r_0^{12} + Q_2^2)$, where $A_2, Q_2$ are constants of integration. Then, the Ricci scalar $R$ given by \eqref{ric-sc} reads
\begin{equation}
R(r) = 3 r^2 \left( 10 A_2 r^{10} - \frac{a^2}{10 D_2^2} \right)   \, .  \label{R2-iii}
\end{equation}
Thus, it follows from Eq.\eqref{fR2-2-iii} that this form of $R$ gives rise to the following $f(r)$
\begin{equation}
f(r) = 40 A_2 D_2 r^9 + \frac{3 a^2}{5 D_2 r}    \, .  \label{f2-iii}
\end{equation}
In order to obtain the the function of $f(R)$, one needs to solve the twelfth degree algebraic equation \eqref{R2-iii}, i.e. the equation $10 A_2 r^{12} - \frac{a^2}{10 D_2^2} r^2 - \frac{R}{3} = 0$, in terms of $r$. After solving this algebraic equation to find at least one real root, one can obtain $f(R)$ by putting this root $r$ into \eqref{f2-iii}.

If we take $P(r)= r^{-6}$ and $N(r) = \left( r_0 / r \right)^6 A(r)$, then the solution of \eqref{P-fR} that gives $f_R$ in terms of $r$ is the same as \eqref{fR2-iii}. Hence, there are two possible cases for this selection. First one is $D_1 \neq 0$ and $D_2 = 0$, and for this condition the obtained solutions from the Eqs.\eqref{sl-1}-\eqref{sl-3} are
\begin{equation}
Q(r) = \frac{a r_0^6 }{15 D_1 } r^{15} +  Q_3  \, ,  \label{Q3-iii}
\end{equation}
\begin{eqnarray}
& & A(r)^2 =   -A_3  +  \frac{\left( 2 a Q_3  -b \right)}{15 D_1 r_0^6}  r^{15}  + \frac{a^2}{ 225 D_1^2} r^{30}  \, , \qquad  \, \label{A3-iii}
\end{eqnarray}
where it should be satisfied the constraint $c = b Q_3 - a ( A_3 r_0^{12} + Q_3^2)$, and $A_3, Q_3$ are integration constants. For the latter solution, the Ricci scalar and the corresponding function $f(r)$ due to the relation \eqref{fR2-2-iii}  are
\begin{eqnarray}
& & R(r) = \frac{240 A_3}{r^2} - \frac{13 a^2}{30 D_1^2} r^{28} \, , \label{R3-iii} \\ & & f(r) = \frac{120 D_1 A_3}{r^4} - \frac{7 a^2}{15 D_1} r^{26} \, . \label{f3-iii}
\end{eqnarray}
Here, it can be obtained the function $f(R)$ from the Eq. \eqref{f3-iii} if the algebraic equation \eqref{R3-iii} in terms of $r$ is solvable. For the second possibility, i.e. if $D_1 = 0$ and $D_2 \neq 0$, we have the following solutions from the first integral relations \eqref{sl-1}-\eqref{sl-3}
\begin{equation}
Q(r) = \frac{a r_0^6 }{16 D_2 } r^{16} +  Q_4  \, ,  \label{Q4-iii}
\end{equation}
\begin{eqnarray}
& & A(r)^2 = - A_4  +  \frac{\left( 2 a Q_4  -b \right)}{16 D_2 r_0^6}  r^{16}  + \frac{a^2}{ 256 D_2^2} r^{32}  \, , \qquad  \, \label{A4-iii}
\end{eqnarray}
where $A_4, Q_4$ are constants of integration and $c = b Q_4 - a ( A_4 r_0^{12} + Q_4^2)$. Thus, the Ricci scalar for this solution yields
\begin{eqnarray}
& & R(r) = \frac{240 A_4}{r^2} - \frac{9 a^2}{16 D_2^2} r^{30} \, , \label{R4-iii}
\end{eqnarray}
which gives the function $f(r)$ from the Eq.\eqref{fR2-2-iii}  as
\begin{eqnarray}
& & f(r) = \frac{96 D_2 A_4}{r^5} - \frac{5 a^2}{8 D_2} r^{27} \, .  \label{f4-iii}
\end{eqnarray}

If we take a bit general expressions $P(r)= r^{\gamma}$ and $N(r) = \left( r / r_0 \right)^{\gamma -1} A(r)$, where $\gamma$ is a real parameter, and use those in the Eq. \eqref{P-fR}, we find $f_R = D_3 \, r^{\gamma}$ in which $D_3$ is a constant of integration. Then, putting those of expressions into the Eq. \eqref{sl-2} and solving it, the angular shift function $Q(r)$ has the form
\begin{equation}
Q(r) = - \frac{a r_0^{1-\gamma}}{3 \gamma D_3} r^{-3 \, \gamma} + Q_5, \label{Q5-iii}
\end{equation}
where $Q_5$ is an integration constant. Considering these in the remaining first integral relations \eqref{sl-1} and \eqref{sl-3}, one can get the following solution
\begin{equation}
A(r)^2 = r^{ 2 ( 1 -3 \gamma)} \left[ - A_5 \, r^{ 6\gamma} +  \frac{\left(b-  2 a Q_5 \right)}{3 \gamma D_3 r_0^{1-\gamma}}  r^{3 \gamma}  + \frac{a^2}{ 9 \gamma^2 D_3^2} \right]  \, , \label{A5-iii}
\end{equation}
where $A_5$ is a constant of integration. Here, the constraint relation $c = b Q_5 - a ( A_5 r_0^{2- 2\gamma} + Q_5^2)$ has to be satisfied. For this solution, the Ricci scalar is
\begin{eqnarray}
& & R(r) = 6 A_5 \gamma^2  - \frac{5 a^2}{6 D_3^2} r^{-6 \, \gamma} \, , \label{R5-iii}
\end{eqnarray}
and the corresponding function $f(r)$ by solving the equation $f' = D_3 r^{\gamma} R'$ is
\begin{eqnarray}
& & f(r) = - \frac{a^2}{D_3} r^{-5 \, \gamma} \, .  \label{f5-iii}
\end{eqnarray}
By using \eqref{R5-iii} and \eqref{f5-iii}, one can easily find $f(R)$ such that
\begin{eqnarray}
& & f(R) = f_1 \left( R_1 - R  \right)^{\frac{5}{6}}  \, ,  \label{f5-2-iii}
\end{eqnarray}
where $f_1 = -(6/5)^{5/6} (a D_3^2)^{1/3}$ and $R_1 = 6 A_5 \, \gamma^2$.

Finally, if we assume $P(r) = r$ and $N(r) = \nu^r A(r)$,  then it follows from the Eq.\eqref{P-fR} that $f_R = D_4\, r \, \nu^r $, where $\nu$ is a real parameter and $D_4$ is an integration constant. Therefore, using the first integral \eqref{sl-2}, the shift function $Q(r)$ can be found as
\begin{equation}
Q(r) = - \frac{a}{3 D_4\, r^3}  + Q_6, \label{Q6-iii}
\end{equation}
with an integration constant $Q_6$. Thus, we can solve the remaining first integrals \eqref{sl-1} and \eqref{sl-3} to find $A(r)^2$ as
\begin{equation}
A(r)^2 = \nu^{ -2 r} \left[ - A_6 \, r^2 +  \frac{\left(b-  2 a Q_6 \right)}{3 D_4 r}  + \frac{a^2}{ 9 D_4^2 \, r^4} \right]  \, , \label{A6-iii}
\end{equation}
together with the constraint $c = b Q_6 - a ( A_6 + Q_6^2)$, where $A_6$ is constant of integration. The Ricci scalar of this solution yields
\begin{eqnarray}
& & R(r) = -\nu^{ -2 r} \Big\{ \ln \nu \left[  4 A_6 \, r +  \frac{\left( 2 a Q_6 - b\right)}{3 D_4 \, r^2}  + \frac{2 a^2}{ 9 D_4^2 \, r^5} \right]  - 6 A_6 + \frac{ 5 a^2}{6 D_4^2 \, r^6} \Big\}  \, , \label{R6-iii}
\end{eqnarray}
which gives rise to
\begin{eqnarray}
& & f(r) = - \frac{\nu^{ -r} }{D_4} \Big\{ \frac{ a^2}{r^5}  + 2 \ln \nu \left[ 4 A_6 D_4^2 \, r^2 +  \frac{ D_4 \left( 2 a Q_6 - b\right)}{ 3 r }  + \frac{2 a^2}{9 r^4} \right] \Big\} \, , \qquad  \label{f6-iii}
\end{eqnarray}
by solving the equation $f' = D_4\, \nu^r \, r \,R'$.

\section{Physical significance of exact solutions}
\label{sec:phys}

The BTZ black hole as the (2+1)-dimensional stationary circularly symmetric solution possesses certain features inherent to (3+1)-dimensional black holes. So, one can naturally expect that the (2+1)-dimensional $f(R)$ gravity may provide new insights towards a better understanding of the physics of  four- and higher dimensional $f(R)$ gravities. Furthermore, it is often much easier to obtain and analyze black hole solutions in three dimensions than in other dimensions.

Defining a curvature stress-energy tensor
\begin{eqnarray}
T_{\mu \nu}^{curv} = \frac{1}{f_R} \left\{ \frac{1}{2} g_{\mu \nu} (f - R f_R) +  (\nabla_{\mu} \nabla_{\nu} - g_{\mu \nu} \square ) f_R \right\}, \quad & &  \label{emt-curv}
\end{eqnarray}
the field equation \eqref{feqs} can be recast in the form:
\begin{equation}
G_{\mu \nu} = R_{\mu \nu} - \frac{1}{2} g_{\mu \nu} R = T_{\mu \nu}^{curv} \, , \label{feqs-curv}
\end{equation}
in absence of the ordinary matter. Then, assuming $f_{_R} \neq 0$, one can rewrite the field equations by using \eqref{feqs-curv} as
\begin{eqnarray}
& & \frac{P' N'}{P N} + \frac{P^2 Q'^2 }{4 N^2} = \frac{\rho_{eff}}{A^2}  \, , \label{feqs1-curv}  \\
& & \frac{N''}{N} + \frac{A' N'}{A N} - \frac{3 P^2 Q'^2 }{4 N^2} = \frac{p_{eff}}{A^2} \, , \label{feqs2-curv} \\ & &  \frac{P''}{P} + \frac{A' P'}{A P} + \frac{P^2 Q'^2 }{4 N^2} = \frac{p_{eff}}{A^2}  + \frac{f'_{_R}}{f_{_R}} ( \frac{N'}{N} - \frac{P'}{P} ) \, , \quad \,\, \label{feqs3-curv}  \\ & &  \frac{Q''}{Q} + \frac{Q'}{Q} \left( \frac{3 P'}{P} + \frac{A'}{A} - \frac{N'}{N} \right) = - \frac{f'_{_R} Q'}{f_{_R} Q}  \, , \label{feqs4-curv}
\end{eqnarray}
where $\rho_{eff}$ and $p_{eff}$ are defined as
\begin{equation}
\rho_{eff} = \frac{A^2}{f_{_R}} \left[ \frac{(f - R f_{_R})}{2 A^2} - f'_{_R} \left( \frac{N'}{N} + \frac{P'}{P} \right)  \right]   \, , \label{rho-eff}
\end{equation}
\begin{equation}
p_{eff} = \frac{A^2}{f_{_R}} \left[ \frac{ (f - R f_{_R})}{2 A^2}  - f'_{_R} \left( \frac{A'}{A} + \frac{N'}{N} \right) - f''_{_R}\right] , \label{p-eff}
\end{equation} \eqref{emt-curv},  \eqref{rho-eff} and \eqref{p-eff}
which are the curvature term contributions to energy density and pressure. Due to the lack of ordinary matter, the latter energy density and pressure may be related with dark matter and dark energy. For the solutions obtained in the latter section, one can construct the equation of state $p_{eff} = w_{_{eff}} \rho_{eff}$, where $w_{_{eff}}$ is the equation of state parameter. When $w_{_{eff}}$ is a constant parameter, one can assume the values $w_{_{eff}} =0$ for pressureless dust,  $w_{_{eff}} = 1$ for stiff matter and $w_{_{eff}} = -1$ for the dark energy. Also, one may consider $w_{_{eff}}$ as a variable equation of state parameter. In (3+1)-dimensions the equation of state $p = \rho/3$, where $w = 1/3$, may be used to describe either actual electromagnetic radiation (photons) or a gas of massless particles in a thermodynamic equilibrium (for example neutrinos). In Ref. \cite{cg2014}, it is represented that it is not appropriate to use the (3+1)-dimensional equation of state parameter for the radiation in (2+1)-dimensions, and they concluded that the equation of state for the radiation in (2+1)-dimensions is $p = \rho/2$.

It is seen from the above construction of the field equations that the non-vanishing components of the curvature stress-energy tensor are
\begin{eqnarray}
& &  T_{rr}^{curv} = \frac{\rho_{eff}}{A^2} \, , \qquad T_{\phi \phi}^{curv} = P^2 p_{eff} \, , \label{emt1-curv} \\ & & T_{\phi t}^{curv} = \frac{1}{2} P^2 Q \left( \frac{p_{eff}}{A^2} + \frac{f'_{_R} Q'}{2 f_{_R} Q} \right) \, , \label{emt2-curv} \\ & & T_{tt}^{curv} =  (P^2 Q^2 - N^2) \rho_{eff} + A^2 N^2 \left( \frac{A'}{A} - \frac{N'}{N} \right) \frac{f'_{_R}}{f_{_R}} + A^2 P^2 Q^2 \left( \frac{P'}{P} - \frac{A'}{A} + \frac{Q'}{Q} \right) \frac{f'_{_R}}{f_{_R}}  \, , \label{emt3-curv}
\end{eqnarray}
in which the components $T_{rr}^{curv}$ and $T_{\phi \phi}^{curv}$, the first term on the right hand side of $T_{\phi t}^{curv}$, and the first two terms on the right hand side of the component $T_{tt}^{curv}$ make up a perfect fluid-like contribution to the curvature stress-energy tensor. We refer the first reference in \cite{viqar} to discussions on how the stress-energy tensor is determined from the metric. Now we are ready to point out what are the physical implications of the exact solutions obtained in previous section. For each solutions, we will set some constants appeared in the solutions to the mass $M$, the angular momentum $J$ and the cosmological constant $\Lambda$ in an appropriate way.

\bigskip

{\bf Case (i):} The form of $f$ in this case is $f(R) = R - 2 \Lambda$. We note here that the choice $d_2 =0$ is necessary, since otherwise the metric \eqref{metric-1} has unphysical asymptotic behaviour \cite{carlip1998}. Then, after setting the remaining constants such that $d_1 = 0, d_3 = \ell^{-2}, k= 2, N_0 = -M$ and $a= J$, the metric \eqref{metric-1} for this case reduces the BTZ-like form
\begin{equation}
ds^2 = - N^2 dt^2 + \frac{P'^2}{N^2} d r^2 + P^2 \left( d \phi  -\frac{J}{2 P^2} dt \right)^2 \, , \label{metric-case-i}
\end{equation}
where $N^2 = -M + \frac{P^2}{\ell^2} + \frac{J^2}{4 P^2}$ and $P = F(r)$. Thus, the Noether constants given by \eqref{c-i-2} becomes $a = J, b= -2 M, c= J / \ell^{2}, \Lambda = -1 / \ell^{2}, I_6 = 0$ and $I_7 = \Lambda$, in which the condition that the cosmological constant must be negative is satisfied. Taking $P(r) =r$ and $P(r)=\frac{\ell}{z} \left( \frac{r}{\ell} \right)^z$ in \eqref{metric-case-i}, we arrive the original rotating BTZ black hole and asymptotically Lifshitz black hole, respectively.
For this solution, the effective energy density $\rho_{eff}$ and the effective pressure $p_{eff}$ can be found by \eqref{rho-eff} and \eqref{p-eff} as
\begin{equation}
p_{eff} = \rho_{eff} = - \Lambda = \frac{1}{\ell^2} \, , \label{p-rho-case-i}
\end{equation}
which satisfies the condition $p_{eff} \geq 0$ and $\rho_{eff} \geq p_{eff}$. It seems that the latter equation can be interpreted as the equation of state for \emph{stiff dark matter}, since $w_{_{eff}} = 1$. The point particle solutions in 2+1 dimensions are good models for parallel
cosmic strings in 3+1 dimensions. An important fact is that the line element \eqref{metric-case-i} with $P(r)=r$, the BTZ black hole, corresponds to the point particle solution which has been also used to draw conclusions about the behavior of cosmic strings \cite{carlip1998}.

In this case, the horizons are located at the positive roots of the equation $N(r)^2 = 0$, i.e. $P^4 - M \ell^2 P^2 + \frac{J^2 \ell^2}{4} = 0$, which has the following exact positive roots
\begin{eqnarray}
& &   P_{\pm} =  \left[ \frac{\ell}{2}\left( M \ell \pm \sqrt{ M^2 \ell^2 - J^2} \right) \right]^{1/2} \, , \label{hor-i-1}
\end{eqnarray}
where $|J| \leq M \ell$. Here, the positive roots given by \eqref{hor-i-1} correspond to the outer and inner horizons ($r_+$ and $r_{-}$) of black hole if $P(r)=r$. For instance, the lapse function $N^2$, the mass $M$ and the angular momentum $J$ for the original BTZ black hole, where $P(r)=r$, can be rewritten in terms of $r_{\pm}$,
\begin{eqnarray}
& & N(r)^2 = \frac{1}{\ell^2 r^2} \left(r^2 - r_+^2 \right)  \left(r^2 - r_{-}^2 \right) \, ,\\ & & M = \frac{r_+^2 + r_{-}^2}{\ell^2} \, , \qquad J = \frac{2 r_+ r_{-}}{\ell} \, . \label{M-J-case-i}
\end{eqnarray}
The angular velocity for the solution \eqref{metric-case-i} is defined by
\begin{equation}
\Omega_H = - \left. \frac{g_{t\phi}}{g_{\phi \phi}} \right|_{r = r_+} =  \frac{J}{2 r_+^2}  \, . \label{ang-vel-i}
\end{equation}
For other forms of $P(r)$, one can easily obtain physically important quantities such as mass, angular momentum and angular velocity in terms of the outer and inner horizons.

\bigskip

{\bf Case (ii):} Here, the function $f(R)$ follows from the Noether symmetry equations as the power law form $f(R)=f_0 R^n$. In this case, we have found five exact solutions for $A=N$ and two ones for $A \neq N$. For physical reason, we set $q_1=0$ throughout this case.

The first solution of this case given by \eqref{ii-sl11} and \eqref{ii-sl12} becomes
\begin{equation}
ds^2 = \frac{4 \, r^{2 \alpha + \beta +1}}{J^2}  d r^2 + r^{\alpha+1} d \phi^2 - \frac{J}{ r^{(\alpha+ \beta)/2} } dt d\phi \, , \label{metric-case-ii-1}
\end{equation}
by taking $a$ as
\begin{equation}
  a = \frac{J}{4} f_0 n R_1 ( 3\alpha + \beta +2) \, ,
\end{equation}
and the parameters $\alpha$ and $\beta$ depend on $n$ as given in \eqref{ii-a-b}. The Ricci scalar of the metric \eqref{metric-case-ii-1} is $R(r)= \frac{n J^2}{32 (n-1)} (\alpha + \beta)(3 \beta -\alpha +4) \, r^{ \frac{\beta +1}{2 (n-1)}}$. It is obviously seen that the above metric is a massless rotating (2+1)-dimensional BTZ-type solution of $f(R) = f_0 R^n$ gravity, where $n\neq 1,1/2, 7/6$. Then, using the definitions of $\alpha$ and $\beta$ given in \eqref{ii-a-b},  the computation of $\rho_{eff}$ and $p_{eff}$ for the metric \eqref{metric-case-ii-1} gives
\begin{eqnarray}
& & p_{eff} = -\frac{J^2(n-1)(2n-1)(4n^2 - 24n +23)}{( 20 n^2-36 n +17)^2 \, r^{ -\frac{4 (2 n-3)}{20 n^2 - 36 n +17}} }  \, , \qquad \label{p-case-ii-1}  \\& & \rho_{eff} = \frac{4 J^2 (n-1)^2 (2n-1)^2 }{ ( 20 n^2-36 n +17)^2 } r^{ \frac{4 (2 n-3)}{20 n^2 - 36 n +17}}  \, , \label{rho-case-ii-1}
\end{eqnarray}
which yields a constant equation of state parameter
\begin{equation}
 w_{_{eff}} = \frac{p_{eff}}{ \rho_{eff} } = - \frac{(4 n^2 - 24 n + 23)}{4 (n-1)(2n-1)}   \, . \label{w-eff-case-ii-1}
\end{equation}
We point out that $w_{_{eff}} = -1$ (\emph{the dark energy}) if $n= -3 /2 \pm \sqrt{7}$, and $w_{_{eff}} = 1/2$ (\emph{the dark radiation}) if $n= 5/4, 5/2$, and $w_{_{eff}} = 1$ ( \emph{the stiff dark matter}) if $n=3/2$. In the solution \eqref{metric-case-ii-1}, one can introduce a special value of $r$, say $r=r_s$, that is a sort of natural unit of length. Of course, this does not make $r=r_s$ an event horizon. Further, the metric \eqref{metric-case-ii-1} has the nonzero angular velocity such that
\begin{equation}
\Omega_H = - \left. \frac{g_{t\phi}}{g_{\phi \phi}} \right|_{r = r_s} =  \frac{J}{2 r_s^{(3 \alpha + \beta + 2)/2}}  \, . \label{ang-vel-ii-1}
\end{equation}

Setting $a= R_1 J$ and $b = \mu R_1 q^2$ which yields that $c=0$ and $k= 2 b = 2 \mu R_1 q^2$, the second solution with the metric coefficients \eqref{ii-sl21} and \eqref{ii-sl22} takes the form
\begin{equation}
  d s^2 = -\frac{\mu q^2}{r} dt^2 + \frac{dr^2}{N^2} + r^2 d\phi^2 -\frac{2 J}{3 r} dt d\phi \, , \label{metric-case-ii-2}
\end{equation}
where $N^2 = \frac{J^2}{9 r^4} + \frac{\mu q^2}{r}$, in which the second term differs from the charge-term of BTZ-Maxwell solutions which is logarithmic form. Here, we can infer from the literature such as \cite{hendi2020} that $q$ is a constant related to the charge of the black hole in power-Maxwell nonlinear electrodynamics, and $\mu$ is a constant parameter to control the character of the charge. So the spacetime \eqref{metric-case-ii-2} is a massless charged BTZ-type black hole with $J \neq 0$. For this solution, the Ricci scalar is $R(r)= -\frac{5 J^2}{6 r^6}$, and $f(R)= f_0 R^{5/6}$. The equation $N^2=0$ has only one root at $r_+$ given by
\begin{equation}
  r_+ = \left( - \frac{J^2}{9 \mu q^2} \right)^{1/3} \, , \label{root-case-ii-2}
\end{equation}
which exists provided $\mu < 0$. Thus the angular velocity of \eqref{metric-case-ii-2} is $\Omega_H = J / (3 r_+^3)$. Also, we find the effective pressure and density for the above solution as follows:
\begin{eqnarray}
& & p_{eff} = \frac{13 J^2}{36 \, r^6} + \frac{\mu q^2}{r^3} \, ,  \qquad \rho_{eff} = \frac{J^2}{36 \, r^6} - \frac{\mu q^2}{2 \, r^3} \, , \quad \label{p-rho-case-ii-2}
\end{eqnarray}
which concludes that the equation of state is
\begin{equation}
  p_{eff} = w_{_{eff}} (r) \rho_{eff} \, , \label{eq-state-case-ii-2}
\end{equation}
where $w_{eff} (r)$ is variable equation of state parameter which has the form:
\begin{equation}
  w_{_{eff}} (r) = \frac{13 J^2 + 36 \mu q^2 r^3}{ J^2 - 18 \mu q^2 r^3} \, . \label{weff-case-ii-2}
\end{equation}
Furthermore, one can deduce from \eqref{p-rho-case-ii-2} the following relations between $p_{eff}$ and $\rho_{eff}$
\begin{equation}
  p_{eff} = -2 \rho_{eff} + \frac{5 J^2}{12 r^6} \, , \qquad p_{eff} = 13 \rho_{eff} + \frac{15 \mu q^2}{2 r^3} \, , \label{eq-state2-case-ii-2}
\end{equation}
which gives that the equation of state parameter is $w_{_{eff}} = -2$ if $J=0, q\neq 0$, and $w_{_{eff}} = 13$ if $q=0, J\neq 0$.

The third solution, where $f(R) = f_0 R^{3/2}$, has the metric coefficients \eqref{ii-sl31} and \eqref{ii-sl32} in which $N=A$, and the spacetime reduces to the well-known BTZ black hole solution by setting $a= R_1 J, b = -2 R_1 M$ and $c= a \ell^{-2}$.

For $n \neq 3/2$, the fourth solution has the following spacetime
\begin{equation}
  ds^2 = M dt^2 + \frac{dr^2}{-M + \frac{J^2}{4 r^2}} + r^2 d\phi^2 - J dt d\phi \, , \label{metric-case-ii-4}
\end{equation}
taking $a = R_1 J, b= -2 R_1 M$ and $k = \frac{2(2n-3)}{2n-1} R_1 M$ in \eqref{ii-sl41}  and \eqref{ii-sl42}. This is a standard (2+1)-dimensional black hole without cosmological constant. For the metric \eqref{metric-case-ii-4}, the Ricci scalar vanishes, i.e. $R=0$, and it has a horizon $r_+ = J /(2 \sqrt{M})$. So it has no extremal limit in the usual sense. In addition, we find for this solution that $p_{eff} = 0$ and $\rho_{eff} = 0$. This means that the spacetime \eqref{metric-case-ii-4} represents true vacuum.

For the $f(R) = f_0 R^{1/2}$ gravity, the fifth solution given by \eqref{ii-sl51} and \eqref{ii-sl52} has the same form and the same properties with metric \eqref{metric-case-ii-4}, but there is no mass parameter, i.e. $M=0$, in this solution.

We obtained the remaining two solutions under the assumption $A \neq N$. Then, setting $b= 0, c=0$ and $N_0 = J K_2 ^{  4(\alpha+1)(n-1) / (2n-3)}$ in \eqref{R-A-case-ii-6}-\eqref{Q2-ii-2},  the sixth solution yields the metric
\begin{equation}
  ds^2 = \frac{dr^2}{A_0^2 r^{ \frac{ 4(n-1) -2\alpha}{2n -1}}} + r^{\alpha +1} d\phi^2 - \frac{2 J  dt d \phi}{ r^{ -\frac{ 4(n-1)(\alpha +1)}{2n-3}}} \, , \label{metric-case-ii-6}
\end{equation}
where $A_0 = - a(2n-3) (J / N _0)^{\frac{2n-3}{2 (2n-1)}} /[ f_0 n (6n-7)(\alpha+1)]$. The Ricci scalar of this solution is
\begin{equation}
  R(r) = \frac{ 8 n (n-1)(6n-7)(\alpha +1)^2 A_0^2}{(1-2n)((2n-3)^2}  r^{-\frac{2(\alpha+1)}{2n-1}} \, ,
\end{equation}
while the form of $f$ is $f(R) = f_0 R^n$. Now, for the energy density and pressure relations, it follows from the Eqs. \eqref{rho-eff} and \eqref{p-eff} that
\begin{eqnarray}
& & p_{eff} = -\frac{ 4 (\alpha+1)^2 (n-1)(n^2 -6n + \frac{23}{4}) A_0^2}{ (2n -1)(2n-3)^2 \,\, r^{ \frac{2 (\alpha+1)}{2n-1}} } \, , \\ & & \rho_{eff} = \frac{ 4 (\alpha+1)^2 (n-1)^2 A_0^2}{ (2n-3)^2} r^{-\frac{2 (\alpha+1)}{2n-1}} \, ,
\end{eqnarray}
where $n \neq 1, 1/2, 3/2$. The latter equations give rise to the same constant equation of state parameter $w_{_{eff}}$ with \eqref{w-eff-case-ii-1}. For the solution \eqref{metric-case-ii-6}, it is not possible to get $w_{_{eff}} = 1$ due to the restriction $n\neq 3/2$.

By setting $a = q_0 J /2, b = - q_0 M$ and $c= a \ell^{-2}$,  the seventh solution in this case includes the metric
\begin{eqnarray}
  ds^2 = - \nu^2 r^{ 2(\nu-1)} A^2 dt^2 + \frac{dr^2}{A^2} + r^{2 \nu} \left( d\phi - \frac{J dt}{2 r^{2 \nu}} \right)^2 , \quad \label{metric-case-ii-7}
\end{eqnarray}
where $$A^2 = -\frac{M}{\nu^2 r^{2(\nu-1)}} + \frac{r^2}{\nu^2 \ell^2} + \frac{J^2}{4 \nu^2 r^{ 2(2\nu-1)}},$$ with $f(R) = f_0 R^{3/2}$. The Ricci scalar of the above metric is a constant, $R = -6 \ell^{-2}$, and the effective pressure and density becomes $p_{eff} = \rho_{eff} = \Lambda$, where $\Lambda = -\ell^{-2}$ is the cosmological constant. It is obviously seen that the metric \eqref{metric-case-ii-7} gives the original BTZ black hole if $\nu =1$. So the metric \eqref{metric-case-ii-7} is more general than the BTZ one, but both this metric and the BTZ spacetime are physically identical. The horizons are given by the condition that the function $A(r)^2 = 0$, and read
\begin{equation}
  r_{\pm} = \left[ \frac{\ell}{2} \left( M \ell \pm \sqrt{M^2 \ell^2 - J^2}\right)  \right]^{\frac{1}{2 \nu}} \, .
\end{equation}
Then, in terms of the horizons $r_{\pm}$, the mass $M$, the angular momentum $J$ and the angular velocity $\Omega_H$ become, respectively
\begin{eqnarray}
& &   M = \frac{ r_+^{2 \nu} + r_{-}^{2 \nu}}{\ell^2} \, , \qquad J = \frac{2 r_+^{\nu} r_{-}^{\nu}}{\ell} \, , \qquad \Omega_H = \frac{r_{-}^{\nu} }{ \ell r_+^{\nu} } \, . \label{MJO}
\end{eqnarray}
In addition, the surface $r_{erg}$ is the surface of infinite
redshift where $g_{tt} (r_{erg})$ vanishes, $r_{erg} = (M \ell^2)^{1/(2 \nu)}$. Obviously, $r_{erg} \geq r_+$. The region $r_+ \leq r \leq r_{erg}$ is called as the ergosphere of the BTZ-type black hole. %

\bigskip

{\bf Case (iii):}  This case implies an arbitrary form of the function $f(R)$. Firstly, for physical reasons, we have to set the constants $Q_i= 0 \, (i=1,...,6)$  appeared in the angular shift function $Q(r)$.

If we set $a= 3 D_1 J, b=-6 D_1 M, c= a \ell^{-2}$ and $A_1= - r_0^{-12} \ell^{-2}$ in the case $D_1 \neq 0$ and $D_2 = 0$, then the metric coefficients $P(r)=r, N(r) = (r /r_0 )^{-6} A(r)$, $Q(r)$ by \eqref{Q1-iii} and $A(r)$ by \eqref{A1-iii} give rise to the spacetime
\begin{eqnarray}
& &   ds^2 = -  \left[ \frac{r^2}{\ell^2}  -  \frac{M r_0^6}{r^4} \right] dt^2 + \frac{dr^2}{A^2} + r^2 d\phi^2 -\frac{J r_0^6}{r^4} dt d\phi \, ,  \label{metric-case-iii-1}
\end{eqnarray}
where $A^2 = \frac{r^{8}}{r_0^{6}} \left[ -M + \frac{1}{\ell^2}\left( \frac{r}{r_0} \right)^6  + \frac{J^2}{4}\left( \frac{r}{r_0} \right)^{-6} \right]$. Then, the Ricci scalar of \eqref{metric-case-iii-1} yields $R =- \frac{30}{\ell^2} (\frac{r}{r_0})^{12} - 6 J^2$, and $f(R) = f_0 ( R + R_0)^{5/6}$ with $R_0 = 6 J^2$. The effective pressure and density can be computed from \eqref{rho-eff} and \eqref{p-eff} such that
\begin{equation}
  p_{eff} = \frac{1}{2} \left[ 4 M \left( \frac{r}{r_0} \right)^{6} - J^2 + \frac{14}{\ell^2} \left( \frac{r}{r_0} \right)^{12} \right] \, , \label{p-eff-case-iii-1}
\end{equation}
and
\begin{equation}
  \rho_{eff} =  2 M \left( \frac{r}{r_0} \right)^{6} + J^2 + \frac{1}{\ell^2} \left( \frac{r}{r_0} \right)^{12}  \, . \label{rho-eff-case-iii-1}
\end{equation}
These can be arranged to give a variable equation of state parameter
\begin{equation}
  w_{_{eff}} (r) =  -\frac{1}{2} \frac{ \left[ J^2 - 4 M\left( \frac{r}{r_0} \right)^6 -\frac{14}{\ell^2} \left( \frac{r}{r_0} \right)^{12} \right]}{ \left[ J^2 + 2 M\left( \frac{r}{r_0} \right)^6 + \frac{1}{\ell^2} \left( \frac{r}{r_0} \right)^{12} \right]}  \, , \label{eqn-state-1}
\end{equation}
which becomes $-1/2$ at the limit $ r \rightarrow 0$.
For the solution \eqref{metric-case-iii-1}, there exist two coordinate singularities corresponding to the outer (event) and inner horizon from $A(r)^2 = 0$,
\begin{equation}
  r_{\pm} =  r_0  \left[ \frac{\ell}{2} \left( M \ell \pm \sqrt{ M^2 \ell^2 - J^2} \right) \right]^{1/6} \, , \label{r-out-in-1}
\end{equation}
where $|J| \leq M \ell$, and $|J| = M \ell$ corresponds to the extreme black hole. The radius of the ergosphere $r_{erg}$ is defined as the stationary limit, which is obtained by solving $g_{tt} = 0$ as follows
\begin{equation}
  r_{erg} = r_0 \left( M \ell^2 \right)^{1/6}  \, . \label{r-erg-1}
\end{equation}
Then, one can express the mass $M$ and the angular momentum (spin) in terms of $r_+$ and $r_{-}$ such as
\begin{equation}
  M = \frac{ r_+^6 + r_{-}^6}{\ell^2 r_0^6} \, , \qquad J = \frac{2 r_{+}^3 r_{-}^3 }{\ell r_0^6} \ .
\end{equation}
The angular velocity of the black hole horizon for metric \eqref{metric-case-iii-1} is given as
\begin{equation}
\Omega_H = - \left. \frac{g_{t\phi}}{g_{\phi \phi}} \right|_{r = r_+} = \frac{J}{2} \left( \frac{r_0}{r_+} \right)^6  \, . \label{ang-vel-1}
\end{equation}

For the second possibility $D_1 = 0$ and $D_2 \neq 0$, we have the spacetime
\begin{eqnarray}
& &   ds^2 = - \left[ \frac{r^2}{\ell^2} - \frac{M r_0^5}{r^3} \right] dt^2 + \frac{dr^2}{A^2} + r^2 d\phi^2 -\frac{J r_0^5}{r^3} dt d\phi \, ,  \label{metric-case-iii-2}
\end{eqnarray}
by taking $a = 5 J D_2 /(2 r_0)$, $b = -5 M D_2 /r_0$ and $A_2 = - 1/(r_0^{12} \ell^{2})$ in the metric coefficients, where
$$A^2 = \frac{r^{4}}{r_0^{2}} \left[ -M \left( \frac{r}{r_0} \right)^5 + \frac{1}{\ell^2}\left( \frac{r}{r_0} \right)^{10}  + \frac{J^2}{4} \right].$$ Here, the Ricci scalar differs from that of \eqref{metric-case-iii-1} as $R = -\frac{30}{\ell^2} (\frac{r}{r_0})^{12} - \frac{15}{8} J^2 (\frac{r}{r_0})^2 $, and we do not have an explicit form of $f(R)$, but we have $f(r) = -\frac{16 a}{ J \ell^2 r_0^{2}} (\frac{r}{r_0} )^9  - \frac{3 a J}{2 \, r_0 \, r}$. For this solution, the $p_{eff}$ and $\rho_{eff}$ are of the form
\begin{equation}
  p_{eff} = -  \frac{r^2}{r_0^2}  \left[ \frac{27 J^2}{16} - 3 M ( \frac{r}{r_0} )^{5} - \frac{7}{\ell^2} ( \frac{r}{r_0} )^{10} \right] \, , \label{p-eff-case-iii-2}
\end{equation}
and
\begin{equation}
  \rho_{eff} =   \frac{r^2}{r_0^2} \left[ \frac{9 J^2}{16}  + \frac{3}{2} M ( \frac{r}{r_0} )^{5} + \frac{1}{\ell^2} ( \frac{r}{r_0} )^{10} \right] \, , \label{rho-eff-case-iii-2}
\end{equation}
which leads to the following variable equation of state parameter
\begin{equation}
  w_{_{eff}} (r) =  - \frac{ \left[ 27 J^2 - 48 M\left( \frac{r}{r_0} \right)^5 -\frac{112}{\ell^2} \left( \frac{r}{r_0} \right)^{10} \right]}{ \left[ 9 J^2 + 24 M\left( \frac{r}{r_0} \right)^5 + \frac{16}{\ell^2} \left( \frac{r}{r_0} \right)^{10} \right]}  \, . \label{eqn-state-2}
\end{equation}
The metric coefficient $A(r)$ of \eqref{metric-case-iii-2} vanishes at $r= r_{\pm}$, where the outer and inner horizons are
\begin{eqnarray}
r_{\pm} = r_0  \left[ \frac{\ell}{2} \left( M \ell \pm \sqrt{ M^2 \ell^2 - J^2} \right) \right]^{1/5} \, (|J| \leq M \ell) \, , & & \quad \label{r-out-in-2}
\end{eqnarray}
The radius of ergosphere $r_{erg}$ by solving $g_{tt}=0$ is obtained as
\begin{equation}
  r_{erg} = r_0 \left( M \ell^2 \right)^{1/5}  \, . \label{r-erg-2}
\end{equation}
Then, the quantities $M$ and $J$ can be written by $r_{\pm}$
\begin{equation}
  M = \frac{ r_+^5 + r_{-}^5}{\ell^2 r_0^5 } \, , \qquad J = \frac{2}{\ell r_0^5} \left( r_{+} r_{-} \right)^{5/2} \ .
\end{equation}
For metric \eqref{metric-case-iii-2}, the angular velocity of the black hole horizon is
\begin{equation}
\Omega_H =  \frac{J}{2} \left( \frac{r_0}{r_+} \right)^5  \, . \label{ang-vel-2}
\end{equation}

The third solution of this case includes $P(r)=r^{-6}$ with the possibility $D_1 \neq 0$ and $D_2 = 0$. Then, assuming $a = - 15 J D_1 /(2 r_0^{21}), b = 15 M D_1 / r_0^{21} $ and $A_3 = -1/(r_0^{12} \ell^{2})$, the relation $N= (r_0 /r)^6 A$, and the metric coefficients \eqref{Q3-iii} and \eqref{A3-iii} bring the spacetime   \eqref{metric-case-iii-1}, \eqref{metric-case-iii-2}, \eqref{metric-case-iii-3}
\begin{eqnarray}
& &  ds^2 = - \left[ \frac{1}{\ell^2 r^{12}}  - \frac{M}{r_0^{12}} \left( \frac{r}{r_0} \right)^{3} \right] dt^2  + \frac{dr^2}{A^2} + \frac{d\phi^2}{r^{12}}  -\frac{J r^3}{r_0^{15}} dt d\phi \, ,  \label{metric-case-iii-3}
\end{eqnarray}
where $A^2 =  -\frac{M}{r_0^{12}} ( \frac{r}{r_0} )^{15} + \frac{1}{\ell^2 r_0^{12}} + \frac{J^2}{4 r_0^{12}} ( \frac{r}{r_0} )^{30}$. Then, the quantities $p_{eff}$ and $\rho_{eff}$ for the above metric are obtained as
\begin{equation}
  p_{eff} = \frac{3}{r_0^{12} r^2} \left[ \frac{14}{\ell^2}  - 4 M ( \frac{r}{r_0} )^{15} + \frac{51 J^2}{16 } ( \frac{r}{r_0} )^{30} \right] \, , \label{p-eff-case-iii-3}
\end{equation}
and
\begin{equation}
  \rho_{eff} =  \frac{9}{r_0^{12} r^2} \left[ \frac{4}{\ell^2}  + M( \frac{r}{r_0} )^{15} + \frac{J^2}{16} ( \frac{r}{r_0} )^{30} \right] \, . \label{rho-eff-case-iii-3}
\end{equation}
Also, the Ricci scalar of this solution is $R = -\frac{240 }{\ell^2 r_0^{12} r^2} - \frac{195 J^2}{8 r_0^{14}} ( \frac{r}{r_0} )^{28}$ and $f(r) = \frac{16 a r_0^9}{J \ell^2 r^4} + \frac{7 a J}{2 r_0^{21}} r ^{26} $.
Then, the black hole solution \eqref{metric-case-iii-3} admits the outer and inner horizons provided $A(r_{\pm}) = 0$ such that
\begin{eqnarray}
r_{\pm} = r_0 \left[ \frac{2 }{\ell J^2} \left( M \ell \pm \sqrt{ M^2 \ell^2 - J^2} \right) \right]^{1/15} \, (|J| \leq M \ell) \, , & & \quad \label{r-out-in-3}
\end{eqnarray}
which implies
\begin{equation}
  M = \frac{ r_0^{15} (r_+^{15} + r_{-}^{15} )}{\ell^2 (r_+ r_{-} )^{15}} \, , \qquad J = \frac{2\, r_0^{15}}{\ell \left( r_{+} r_{-} \right)^{15/2} }  \ .
\end{equation}
The radius of ergosphere and the angular velocity of the event horizon then become
\begin{equation}
  r_{erg} = \frac{r_0}{\left( M \ell^2 \right)^{1/15} } \, , \qquad \Omega_H = \frac{J}{2} \left( \frac{r_+}{r_0}\right)^{15} \, .
\end{equation}

If we consider $D_1 = 0$ and $D_2 \neq 0$ with $P(r)=r^{-6}$, and we set $a= - 16 J D_2 /(2 r_0^{22}), b= 16 M D_2 r_0^{22}$ and $A_4 = -r_0^{-12}\ell^{-2}$ in \eqref{Q4-iii} and \eqref{A4-iii}, then the fourth solution occurs as
\begin{eqnarray}
& &   ds^2 = - \left[ \frac{1}{\ell^2 r^{12}} - \frac{M}{r_0^{12}} \left( \frac{r}{r_0} \right)^{4} \right] dt^2 + \frac{dr^2}{A^2} + \frac{d\phi^2}{r^{12}}  -\frac{J r^4}{r_0^{16}} dt d\phi \, ,  \label{metric-case-iii-4}
\end{eqnarray}
in which $A^2 =  - \frac{M}{r_0^{12}} ( \frac{r}{r_0} )^{16} + \frac{1}{\ell^2 r_0^{12} }  + \frac{J^2}{4 r_0^{12}} ( \frac{r}{r_0} )^{32}$. Here the Ricci scalar becomes $R = -\frac{240 }{\ell^2 r_0^{12} r^2} - \frac{36 J^2}{ r_0^{14}} ( \frac{r}{r_0} )^{30}$ and the function $f(r)$ has the form $f(r) = \frac{12 a r_0^{10}}{J \ell^2 r^5} + 5 a J r_0^5 ( \frac{r}{r_0})^{27} $. Furthermore, the latter solution has the effective pressure $p_{eff}$ and the effective density $\rho_{eff}$ in the form
\begin{equation}
  p_{eff} = \frac{1}{r_0^{12} r^2} \left[ \frac{42}{\ell^2}  - 18 M ( \frac{r}{r_0} )^{16} + \frac{29 J^2}{2} ( \frac{r}{r_0} )^{32} \right] \, , \label{p-eff-case-iii-4}
\end{equation}
and
\begin{equation}
  \rho_{eff} =  \frac{1}{r_0^{12} r^2} \left[ \frac{36}{\ell^2} + 12 M ( \frac{r}{r_0} )^{16} + J^2 ( \frac{r}{r_0} )^{32} \right] \, . \label{rho-eff-case-iii-4}
\end{equation}
Now, it is easy to find exact expressions for the roots of the $A(r) = 0$ which are the horizons of the metric \eqref{metric-case-iii-4} such as
\begin{eqnarray}
r_{\pm} = r_0 \left[ \frac{2 }{\ell J^2} \left( M \ell \pm \sqrt{ M^2 \ell^2 - J^2} \right) \right]^{1/16} \, , & & \quad \label{r-out-in-4}
\end{eqnarray}
where $|J| \leq M \ell$, and $|J| = M \ell$ for the extreme black hole. Thus, one can obtain $M$ and $J$ by using \eqref{r-out-in-4} as
\begin{equation}
  M = \frac{ r_0^{16} (r_+^{16} + r_{-}^{16} )}{\ell^2 (r_+ r_{-} )^{16}} \, , \quad J = \frac{2 r_0^{16}}{\ell (r_{+} r_{-})^8 } \ .
\end{equation}
For the metric \eqref{metric-case-iii-4}, the quantities $r_{erg}$ and $\Omega_H$ are
\begin{equation}
  r_{erg} = \frac{r_0}{ \left( M \ell^2 \right)^{1/16} }\, , \quad \Omega_H = \frac{J}{2} \left( \frac{r_+}{r_0}\right)^{16} \, .
\end{equation}

The fifth solution expressed by the metric coefficients $P = r^{\gamma}$, $N = (r /r_0)^{\gamma-1} A$, $Q$ by \eqref{Q5-iii} and $A$ by \eqref{A5-iii} can be written as \begin{eqnarray}
& &  ds^2 = - \left( \frac{r}{r_0} \right)^{2 (\gamma-1)} A^2 dt^2  + \frac{dr^2}{A^2} + r^{2 \gamma} \left[ d\phi - \frac{J}{2} r^{-3\gamma} dt \right]^2 \, , \label{metric-case-iii-5}
\end{eqnarray}
when we set $a= \frac{3}{2} \gamma J D_3 r_0^{\gamma -1}, b = -3 \gamma D_3 M r_0^{-4\gamma -1}$ and $A_5 = -r_0^{-2( 4\gamma +1)} \ell^{-2}$, where we have $$A^2 = r_0^{2 \gamma}  \left(\frac{r}{r_0} \right)^2 \left[ \frac{1}{\ell^2 r_0^{10\gamma} }- M r_0^{\gamma} r^{-3\gamma} + \frac{J^2}{4} r^{-6\gamma} \right].$$ Then, a straightforward calculation of $p_{eff}$ and $\rho_{eff}$ gives
\begin{equation}
   p_{eff} = \frac{\gamma^2}{r_0^2} \left[ \frac{r_0^{-8\gamma}}{\ell^2} - M ( \frac{r}{r_0} )^{-3 \gamma} + \frac{13 J^2}{16 r_0^{-2\gamma}} r^{-6 \gamma } \right] \, , \label{p-eff-case-iii-5}
\end{equation}
and
\begin{eqnarray}
   \rho_{eff} = \frac{\gamma^2}{r_0^2} \left[  \frac{ r_0^{-8\gamma} }{\ell^2} + \frac{M}{2} ( \frac{r}{r_0} )^{-3 \gamma} + \frac{ J^2}{16 r_0^{-2\gamma}} r^{-6 \gamma } \right]. \,\, & &  \label{rho-eff-case-iii-5}
\end{eqnarray}
The Ricci scalar can be recast in the form $R= -\frac{6}{\ell^2}\gamma^2 r_0^{-2(4 \gamma +1)} - \frac{15}{8} \gamma^2 J^2 r_0^{2(\gamma-1)} r^{-6\gamma}$, and the functions $f(r)$ and $f(R)$ become $f(r)= -\frac{3}{2} a \gamma J r_0^{\gamma-1} r^{-5\gamma}$ and $f(R)= f_1 ( R_1-R)^{5/6}$ with $R_1 = -6 \gamma^2 \ell^{-2} r_0^{-2(4 \gamma +1)}$, respectively.
For the metric \eqref{metric-case-iii-5}, the existence of horizons requires the vanishing of the $g^{rr}$ component, i.e. $A(r)^2 = 0$. The roots of the latter equation give rise to the outer and inner horizons such as
\begin{eqnarray}
r_{\pm} = r_0^{5/3} \left[ \frac{\ell}{2} \left( M \ell \pm \sqrt{ M^2 \ell^2 - J^2} \right) \right]^{ \frac{1}{3 \gamma} } \, , & & \quad \label{r-out-in-5}
\end{eqnarray}
which yields that
\begin{equation}
  M = \frac{  r_{+}^{3\gamma} + r_{-}^{-3\gamma} }{\ell^2 r_0^{10 \gamma} }\, , \qquad J = \frac{2 }{\ell r_0^{5 \gamma} } (r_{+} r_{-})^{ 3 \gamma /2}  \ .
\end{equation}
Finally, the radius of ergosphere and the angular velocity of the event horizon are
\begin{equation}
  r_{erg} = r_0^{5/3} \left( M \ell^2 \right)^{ \frac{1}{3 \gamma}} \, , \qquad \Omega_H = \frac{J}{2} r_+^{-3 \gamma} \, .
\end{equation}

By setting $a= 3 J D_4 /2, b= -3 D_4 M$ and $A_6 = -\ell^{-2}$, the resulting spacetime for the sixth solution of this case is
\begin{equation}
  ds^2 = -\left( \frac{r^2}{\ell^2} - \frac{M}{r} \right) dt^2 + \frac{dr^2}{A^2} + r^2 d\phi^2 -\frac{J}{r} dt d\phi \, , \label{metric-case-iii-6}
\end{equation}
where we have used $P(r)=r, N(r) = \nu^r A(r)$, $Q(r)$ by \eqref{Q6-iii} and $A(r)$ by \eqref{A6-iii} together with $$A^2 = \nu^{-2 r} \left( - \frac{M}{r} + \frac{r^2}{\ell^2 } + \frac{J^2}{4 r^4} \right).$$ The corresponding Ricci scalar for metric \eqref{metric-case-iii-6} is
\begin{eqnarray}
& & R = \nu^{-2r} \left[ \ln \nu \left( \frac{M}{r^2} - \frac{4 r}{\ell^2} + \frac{J^2}{2 r^5} \right) + \frac{6}{\ell^2} + \frac{15 J^2}{8 r^6} \right] \, , \nonumber
\end{eqnarray}
and the function $f(r)$ yields
\begin{eqnarray}
& & f(r) = -\frac{3}{2} a J \nu^{-r} \left[  \frac{1}{r^5} + \frac{4}{9} \ln \nu  \left( \frac{2 M}{J^2 r}  - \frac{8 r^2}{\ell^2 J^2} + \frac{1}{r^4} \right) \right] \, . \nonumber
\end{eqnarray}
For the solution \eqref{metric-case-iii-6}, $p_{eff}$ and $\rho_{eff}$ become
\begin{eqnarray}
& &   p_{eff} = \nu^{-2 r} \Big[ - \frac{M}{r^3}  + \frac{1}{\ell^2} + \frac{13 J^2}{16 r^6} + \frac{\ln \nu}{2} \left( \frac{J^2}{r^5}  -\frac{M}{r^2} -\frac{2 r}{\ell^2} \right) \Big] \, , \label{p-eff-case-iii-6}
\end{eqnarray}
and
\begin{eqnarray}
   \rho_{eff} = \nu^{-2 r} \left[ \frac{1}{\ell^2}  + \frac{M}{2 r^3} + \frac{ J^2}{16 r^6} \right]. \,\, & &  \label{rho-eff-case-iii-6}
\end{eqnarray}
The outer and inner horizons which are the black hole horizons, concerning the positive mass black hole spectrum with spin ($J\neq 0$) of the line element \eqref{metric-case-iii-6} are given by
\begin{eqnarray}
r_{\pm} = \left[ \frac{\ell}{2} \left( M \ell \pm \sqrt{ M^2 \ell^2 - J^2} \right) \right]^{ \frac{1}{3}} \, , & & \quad \label{r-out-in-6}
\end{eqnarray}
and therefore, in terms of the inner and outer horizons, the black hole mass and the angular momentum are given, respectively, by
\begin{equation}
  M = \frac{  r_{+}^{3} + r_{-}^{3}}{\ell^2 } \, , \qquad J = \frac{2 }{\ell } (r_{+} r_{-})^{3/2}  \ .
\end{equation}
The ergosphere radius and angular velocity $\Omega_H$ of the event (outer) horizon can be computed
\begin{equation}
  r_{erg} = \left( M \ell^2 \right)^{ \frac{1}{3}} \, , \qquad \Omega_H = \frac{J}{2 r_+^3} \, .
\end{equation}

We conclude for this case that the equation of state parameter $w_{_{eff}} = p_{eff} / \rho_{eff}$ obtained for each of the solution is a function of $r$.

\subsection{Thermodynamics}
\label{subsec:thermo}

In the previous part of this section, we have studied the properties of obtained solutions associated with the black hole event horizon $r_+$, the effective energy density and pressure. In this subsection, we investigate the thermodynamic analysis of the solutions throughout this study.
For three-dimensional rotating black hole metric \eqref{metric-1} in the framework of $f(R)$ gravity, the derivation of thermodynamic quantities such as temperature and entropy for each of the obtained BTZ-type solutions in this study can be accomplished. It is worth pointing out that the entropy of BTZ black holes, $S = 4 \pi r_+$, does not necessarily holds in the case of extended theories of gravity. In $f(R)$ theory of gravity the horizon entropy of the black hole has the following formula \cite{akbar2007}
\begin{equation}
S = \left. \frac{A_h \, f_R}{4 G_3} \right|_{r=r_+} \, , \label{entropy-1}
\end{equation}
where $A_h$ is the horizon area of black holes, in (2+1)-dimensions a circumference, $A_h = 2 \pi r_+$, $G_3$ being the three-dimensional gravitational constant, and $r_+$ is the radius of the event horizon of the black hole. Here the units are such that $G_3 = 1/8$.
The thermodynamic quantities for the obtained solutions should satisfy the first law of thermodynamics
\begin{equation}
  d M = T d S + \Omega_H d J \, , \label{thermo}
\end{equation}
where the Hawking temperature $T$ and angular velocity $\Omega$ are given by
\begin{eqnarray}
& &   T = \left. \frac{\partial M}{\partial S} \right|_{J, \ell} \, ,  \qquad  \Omega_H = \left. \frac{\partial M}{\partial J} \right|_{S, \ell} \, . \label{temp}
\end{eqnarray}
Also, in terms of the formula
\begin{equation}
  C_J =  \left.  \frac{\partial M}{\partial T} \right|_{J, \ell} = \left. T \frac{\partial S}{\partial T} \right|_{J, \ell} \, ,  \label{heat-cap}
\end{equation}
one can get the heat capacity of the hole which determines the thermodynamic stability. The black hole is locally stable if $C_J \geq 0$, while the corresponding black hole is locally unstable if $C_J < 0$.
Now, first we find the Hawking temperature of the BTZ-type black holes obtained in cases (i)-(iii) and then check that the first law of thermodynamics \eqref{thermo} is satisfied for these solutions. Smarr relation \cite{smarr}, together with the first law of black hole thermodynamics, has a main role in black hole physics. Furthermore, we can get the Smarr-type mass formula of the obtained BTZ-type black holes, and these Smarr-type relations may be sometimes useful in the Euclidean approach to quantum gravity. Now we investigate the thermodynamics for each BTZ-type black holes given in cases (i)-(iii).

\bigskip

{\bf Case (i):} Making use of Eq. \eqref{entropy-1} and the relations $M$ and $J$ written by $r_{\pm}$, one can get the mass formula in terms of $S, J$ and $\ell$ after selecting the metric coefficient $P$. For this case, the entropy of the BTZ-type black hole \eqref{metric-case-i} is $S= 4 \pi r_+$, since $f(R) = R -\Lambda$, that is $f_R =1$. The mass $M$ and angular momentum $J$ for the BTZ black hole in terms of $r_{\pm}$ are already given by \eqref{M-J-case-i}, and for the Lifshitz black hole these quantities are of the form
\begin{equation}
  M = \frac{r_{+}^{2 z} + r_{-}^{2 z}}{z^2 \ell^{2 z}} \, , \qquad J = \frac{2 r_{+}^z r_{-}^z}{z^2 \ell^{2 z}} \, . \label{M-J-case-i-2}
\end{equation}
Then, it follows immediately that for the BTZ black hole and the Lifshitz black hole the mass formula of the spacetime \eqref{metric-case-i} reads, respectively
\begin{equation}
  M = \frac{S^2}{16 \pi^2 \ell^2} + \frac{4 \pi^2 J^2}{S^2} \quad {\rm for} \quad P(r) = r \, , \label{M-case-i-1}
\end{equation}
and
\begin{equation}
  M= \frac{S^{2 z}}{ z^2 (4 \pi \ell)^{2 z}} + \frac{z^2 (4\pi \ell)^{2 z}  J^2 }{4 \ell^{2} S^{2 z}} \,\, {\rm for} \,\, P(r) = \frac{\ell}{z} (\frac{r}{\ell})^z . \label{M-case-i-2}
\end{equation}
Now, using \eqref{M-case-i-1} and \eqref{M-case-i-2}, the Hawking temperatures and angular velocities are obtained from Eq. \eqref{temp} as
\begin{equation} \label{T-case-i-1}
T =\left\{
\begin{array}{ll}
 \frac{ r_+^{2 } - r_{-}^{2 } }{2 \pi \ell^{2} r_+}  \quad \,\,\, {\rm for} & P(r) = r,   \\
 \frac{ r_+^{2 z} - r_{-}^{2 z} }{2 \pi z \ell^{2 z} r_+} \quad {\rm for}  & P(r) = \frac{\ell}{z} (\frac{r}{\ell})^z \, ,
\end{array}
\right.
\end{equation}
and
\begin{equation} \label{Omega-case-i-1}
\Omega_H =\left\{
\begin{array}{ll}
 \frac{ r_{-} }{ \ell r_+}  \quad {\rm for} & P(r) = r,   \\
 \frac{ r_{-}^{ z} }{  \ell r_+^z} \quad {\rm for}  & P(r) = \frac{\ell}{z} (\frac{r}{\ell})^z.
\end{array}
\right.
\end{equation}
The heat capacities at constant angular momentum can be computed as
\begin{equation} \label{heatcap-case-i-1}
C_J =\left\{
\begin{array}{ll}
 \frac{ 4 \pi r_{+} \Delta }{ 2 - \Delta}   & {\rm for} \,\, P(r) = r,   \\
 \frac{ 8 \pi r_{+} z^3 \Delta }{C_1 + C_2 \Delta }   & {\rm for} \,\,  P(r) = \frac{\ell}{z} (\frac{r}{\ell})^z \, ,
\end{array}
\right.
\end{equation}
where $C_1 = 2 z^4 - z^3 + 2 z + 1$, $C_2 = 2 z^4 - z^3 - 2 z -1$ and $\Delta = [1 - J^2 /(M \ell)^2 ]^{1/2}$. The first relation in  \eqref{heatcap-case-i-1} is the usual heat capacity of BTZ black holes \cite{clz}. Also, the quantities $T, S, \Omega_H$ and $J$ obtained in the above give rise to the Smarr-like mass formulas
\begin{equation} \label{mass-forml-case-i-1}
M =\left\{
\begin{array}{ll}
 \frac{1}{ 2 } T S + \Omega_H J   & {\rm for} \,\, P(r) = r,   \\
 \frac{1}{2 z} T S + \Omega_H J  & {\rm for} \,\, P(r) = \frac{\ell}{z} (\frac{r}{\ell})^z \, .
\end{array}
\right.
\end{equation}
Thus, varying the above mass formulas yield the conventional differential expression of the first law of black hole thermodynamics given in \eqref{thermo}.

\bigskip

{\bf Case (ii):} In this case $f(R) = f_0 R^n$, and so $f_R = n f_0 R^{n-1}$.
Firstly, we point out that the solutions \eqref{metric-case-ii-1} and \eqref{metric-case-ii-6} are a massless BTZ-type black holes, they have zero Hawking temperature, zero entropy, and vanishing heat capacity. These are the same as the corresponding quantities of usual extremal BTZ black holes. The spacetime \eqref{metric-case-ii-2} is also a massless but charged BTZ-type black hole with spin ($J \neq 0$). This has zero Hawking temperature, vanishing heat capacity, but a constant entropy such as $S = 4 \pi f_1$ where $f_R = f_1 /r_+$. The metric \eqref{metric-case-ii-4} is a standard (2+1)-dimensional black hole without cosmological constant.

For the black hole solution \eqref{metric-case-ii-7}, the mass $M$, angular momentum $J$ and angular velocity $\Omega_H$ parameters are given in \eqref{MJO}. Therefore, by expressing the mass formula in terms of $S$ and $J$
\begin{equation}
  M = \frac{S^{2 \nu}}{\ell^2 f_2^{2 \nu}} + \frac{ (f_2 \ell J)^{2\nu}}{(4 \ell)^{\nu} S^{2 \nu}} \, ,
\end{equation}
and using the entropy relation $S= f_2 r_+$, one can compute the Hawking temperature as
\begin{equation}
  T = \frac{2 \nu ( r_+^{2 \nu} - r_{-}^{2 \nu}) }{\ell^2 f_2 r_+} \, ,
\end{equation}
where $f_2 = 6 \pi f_0 \sqrt{6 \Lambda}$. Thus, the Smarr-like formula of this solution becomes
\begin{equation}
  M = \frac{1}{2 \nu} T S + \Omega_H J \, ,
\end{equation}
which easily verifies the mass differential \eqref{thermo}. By considering the relation \eqref{heat-cap}, we obtain the heat capacity for BTZ-type black hole \eqref{metric-case-ii-7} in the form
\begin{equation}
  C_J = \frac{f_2 r_+ \Delta}{\nu ( 2 \nu - \Delta)} \, .
\end{equation}

\bigskip

{\bf Case (iii):} In this case, it appears that there are six new BTZ-type black hole solutions. For these solutions, we calculated analytic expressions for thermodynamic quantities $S, T$ and $C_J$, and the corresponding Smarr-like formulas, which are given in Table \ref{Tab1}. Note that the thermodynamic quantities $T, S, J$ and $M$  for each black holes in this case obey the first law of thermodynamics \eqref{thermo}.


\begin{table*}
\caption{\label{Tab1} List of thermodynamic quantities the entropy $S$, the temperature $T$ and the heat capacity $C_J$ together with the Smarr-like formulas for BTZ-type black hole solutions in case (iii) are presented. }
\begin{ruledtabular}
\begin{tabular}{ccccc}
Solution & $S$ & $T$ & $C_J$ & Smarr-like formula \\
\hline
Eq. \eqref{metric-case-iii-1} & $\frac{4 \pi D_1}{r_+}$ & $  \frac{3 r_+ ( r_{-}^6 - r_+^6)}{2 \pi D_1 \ell^2 r_0^6} $  & $ - \frac{ 4 \pi D_1 \Delta}{ r_+ ( 6 + \Delta)} $ with $D_1 = \frac{a}{3 J}$  & $M= -\frac{1}{6} T S + \Omega_H J$ \\\\
Eq. \eqref{metric-case-iii-2} & $\frac{4 \pi D_2}{r_+^2}$ & $ \frac{5 r_+^2 ( r_{-}^5 - r_+^5)}{8 \pi D_2 \ell^2 r_0^{10}} $  & $ - \frac{ 16 \pi D_2 r_0^5 \Delta}{ r_+^2 ( 5 + 2 \Delta)} $ with $D_2 = \frac{2 a r_0}{5 J} $ & $M= -\frac{2}{5} T S + \Omega_H J$ \\\\
Eq. \eqref{metric-case-iii-3} & $\frac{4 \pi D_1}{r_+}$ & $ \frac{15 r_0^{15} r_+}{4 \pi D_1 \ell^2} \left( \frac{1}{r_+^{15}} - \frac{1}{r_{-}^{15}} \right) $  & $ - \frac{ 4 \pi D_1 \Delta}{ r_+ ( 15 + \Delta)} $ with $D_1 = - \frac{2 a r_0^{11}}{15 J}$  & $M= \frac{1}{15} T S + \Omega_H J$ \\\\
Eq. \eqref{metric-case-iii-4} & $\frac{4 \pi D_2}{r_+^2}$ & $ \frac{2 r_0^{16} r_+^2 }{\pi D_2 \ell^2} \left( \frac{1}{r_+^{16}} - \frac{1}{r_{-}^{16}} \right) $ &  $ - \frac{ 4 \pi D_2 \Delta}{ r_+^2 ( 8 + \Delta)} $ with $D_2 = - \frac{a r_0^{22}}{8 J}$  & $M= \frac{1}{8} T S + \Omega_H J$ \\\\
Eq. \eqref{metric-case-iii-5} & $4 \pi D_3 r_+^{\gamma + 1}$ & $ \frac{ 3 \gamma r_+^{-(\gamma+1)} ( r_+^{3 \gamma} - r_{-}^{3 \gamma})}{4 \pi D_3 \ell^2 ( 1+ \gamma) r_0^{10 \gamma}} $  & $ \frac{4 \pi D_3 r_+^{\gamma+1} \Delta}{\frac{3 \gamma}{\gamma+1} - \Delta} $  with $D_3 = \frac{2 a r_0^{1-\gamma}}{3 \gamma J}$  & $M= \frac{(\gamma +1)}{3 \gamma} T S + \Omega_H J$ \\\\
Eq. \eqref{metric-case-iii-6} & $4 \pi D_3 r_+^2 \nu^{r_+}$ & $ \frac{3 \nu^{r_+} ( r_{-}^3 - r_+^3)}{4 \pi \ell^2 r_+^2} $  & $ \frac{4 \pi D_4 r_+^2 \nu^{r_+} \Delta}{ 3 D_4 - (3 + 2 D_4) \Delta} $ with $D_4 = \frac{2 a}{3 J}$ & $M= -\frac{1}{3} T S + \Omega_H J$ \\
\end{tabular}
\end{ruledtabular}
\end{table*}

\section{Conclusions}
\label{sec:conc}

In this study, we have derived the Noether symmetries of a canonical Lagrangian for $f(R)$ theory of gravity in background of three dimensional rotating black hole spacetime \eqref{metric-1}. Using the effective point-like Lagrangian \eqref{lagr-fr} of this spacetime in terms of its configuration space variables $N, A, Q$ and $P$ which are the metric coefficients, and $R$ (the Ricci scalar), and their velocities $N', A', Q', P'$ and $R'$, we have determined the kinetic metric $\sigma_{i j}$ by \eqref{km-fr} in the configuration space of the system. Thus we have considered this kinetic metric and used it to calculate and classify Noether symmetry generators by the derived geometrical Noether symmetry conditions \eqref{ngs-eq}. Later, we obtained the first integrals for each of the Noether symmetries admitted by Lagrangian of representing physical system. Furthermore, we have used the first integrals of motion in order to generate new exact solutions for the $f(R)$ gravity theory of three-dimensional rotating black hole metric \eqref{metric-1}. Also, we have worked the physical properties of these new exact solutions in the previous section. We would like to stress that our results are richer than the strict Noether symmetry approach because we have considered the Noether symmetry approach with a gauge term which also includes the term $\xi \partial_r$ in the generator.

The Noether symmetry approach considered in this work is capable to construct exact solutions of field equations for any gravity theory by reducing their complexity through the first integral(s) of motion \cite{tp2011a,camci2012,camci2016,Bahamonde:2016} without using the cyclic variables. In order to find out analytical solutions of field equations for the $f(R)$ gravity in three dimensional BTZ-like black hole spacetime \eqref{metric-1}, the obtained Noether first integrals have mainly been considered in the cases (i), (ii) and (iii). Also, the equation \eqref{P-fR} has played a key role of finding new exact black hole solutions of (2+1)-dimensional $f(R)$ theory of gravity. Throughout the paper we have denoted  the constants of motion $I_1, I_2$ and $I_3$ by $a, b$ and $c$, respectively, which are valid for any form of $f(R)$. Also, $I_5$ is represented by $k$ in the cases (i) and (ii). We have firstly considered the case (i) where there are \emph{seven} Noether symmetries and the form of $f$ is $f(R)= R - 3 \Lambda$. Using the rearranged first integrals in this case, we have found a general solution of the metric coefficients, which are \eqref{i-PQ} and \eqref{i-N2}, depending on a function $F(r)$ defined by \eqref{def-F}. We have concluded the case (i) representing two examples, the well-known BTZ black hole solution and asymptotically Lifshitz black hole solution. The obtained metric functions \eqref{i-PQ} and \eqref{i-N2} of this case are very generic to produce any other black hole solutions. The most important finding in the case (i) is the fact that the mass and angular momentum of black hole, and the cosmological constant are Noether constants for the well-known BTZ black hole. In the case (ii), we have found \emph{five} Noether symmetries with the power law form $f(R) = f_0 R^n$ from the Noether symmetry equations \eqref{ngs-eq}. Then, starting from equation \eqref{P-fR} which relates the metric functions $P, A, N$ and the Ricci scalar $R$ under the assumption either  $A=N$ or $A \neq N$, we have obtained the metric functions from the first integral equations \eqref{sl-1} and \eqref{ii-eq1}-\eqref{ii-eq4}, which are solutions of the $f(R)$ gravity. In this case, we have found five different solutions for $A=N$ and two different ones for $A \neq N$. In the case (iii), it is found from the Noether symmetry equations that there are \emph{four} Noether symmetries for an arbitrary form of the function $f(R)$. As a first example of this case, we have considered the ansatz $R(r) = K_0 r^m$ proposed in \cite{darabi2013} which yields the explicit form of the function $f(R)$ as in \eqref{fR1}. Unfortunately, this selection of the Ricci scalar $R(r)$ does not provide the metric functions satisfying all the field equations as explained in the first part of case (iii). Afterwards, for $A \neq N$, we have obtained some Lifshitz-like new solutions taking $P(r)= r$ and $N(r) = ( r/ r_0)^z A(r)$, \emph{or} $P(r)= r^{-6}$ and $N(r) = ( r/ r_0)^6 A(r)$,  \emph{or} $P(r)= r^{\gamma}$ and $N(r) = ( r/ r_0)^{\gamma -1} A(r)$, \emph{or} $P(r)= r$ and $N(r) = \nu^r A(r)$, which brings the $f_R$ in terms of $r$ from the equation \eqref{P-fR}. In some solutions, we are able to find the function $f(R)$ explicitly, but in the other ones we have only found $f(r)$ due to the difficulty of solving the algebraic equations of high degree.

By transforming the field equations \eqref{feqs} to the usual form \eqref{feqs-curv}, we have introduced a curvature stress-energy tensor \eqref{emt-curv}, and defined energy density $\rho_{_{eff}}$ and pressure $p_{_{eff}}$ as curvature term contributions by \eqref{rho-eff} and \eqref{p-eff}, respectively. For each of the solutions given in cases (i)-(iii), we found the effective pressure $p_{_{eff}}$ and energy density $\rho_{_{eff}}$, and so the corresponding effective equation of state parameter $w_{_{eff}}$ which is a constant or a variable one. In case (i), the BTZ-like solution \eqref{metric-case-i} has a constant equation of state parameter $w_{_{eff}} = 1$, a stiff dark matter. In case (ii), the solution \eqref{metric-case-iii-2} has only a variable equation of state parameter given in \eqref{eqn-state-2}, and the remaining ones have a constant equation of state parameters. For case (iii), all of the solutions which are \eqref{metric-case-iii-1}, \eqref{metric-case-iii-2}, \eqref{metric-case-iii-3}, \eqref{metric-case-iii-4}, \eqref{metric-case-iii-5} and \eqref{metric-case-iii-6},  have the variable equation of state parameters. Through the section \ref{sec:phys}, we have determined the mass $M$, the angular momentum $J$ and the angular velocity $\Omega_H$ in terms of the event horizon $r_+$ and the inner horizon $r_{-}$ for all the solutions. Then, using the functional form of the mass $M(r_+,r_-)$ and the angular momentum $J(r_+,r_-)$, we have determined the thermodynamic quantities such that the Hawking temperature, the entropy and the heat capacity in the cases (i)-(iii) for the BTZ-like black hole solutions. We have shown that all the obtained solutions satisfy the first law of thermodynamics, and also attained the Smarr-like mass formulas of the new BTZ-type black holes.



In the Appendix A, we have solved the metric symmetries for the kinetic metric $\sigma_{i j}$ of the configuration space given by \eqref{km-fr}. It is found in this appendix that for any form of $f(R)$ the kinetic metric admits at least 6-dimensional Killing algebra, 7-dimensional homothetic algebra, and 7-dimensional conformal Killing algebra if the function $\Phi$ depends only on $A$. For the linear form of $f(R) = R - 2 \Lambda$, we have explored that the kinetic metric admits a 10-dimensional Killing algebra, 11-dimensional homothetic algebra and 11-dimensional conformal Killing algebra if $\Psi =\Psi (A)$. Thus, we have represented that some of the Noether symmetries are the metric symmetries of the kinetic metric $\sigma_{i j}$.

\section*{Appendix A: Spacetime symmetry}
\label{app-sym}

A conformal Killing vector (CKV) ${\bf Y}$ have to satisfy
\begin{equation}
\pounds_{\bf Y} g_{\mu \nu} = 2 \psi (x^{\alpha}) g_{\mu \nu} \, ,
\end{equation}
where $g_{\mu \nu}$ is the metric tensor,  $\pounds_{\bf Y}$ is the Lie derivative operator along ${\bf Y}$ and $\psi (x^{\alpha})$ is a conformal factor. When $\psi_{;\mu \nu} \neq 0$, the CKV field is said to be {\it proper} \cite{katzin}. The vector field ${\bf Y}$ is called the special conformal Killing vector (SCKV) field if $\psi_{;\mu \nu} = 0$, the homothetic vector (HV) field if $\psi_{, \, \mu} = 0$, e.g. $\psi$ is a constant, and the Killing vector (KV) field which gives the isometry if $\psi =0$.
The metric \eqref{metric-1} is stationary and axially symmetric, with the KVs $\partial_t$ and $\partial_{\phi}$ which are describe the two parameters, mass $M$ and angular momentum
(spin) $J$, respectively. The BTZ black hole solution in (2 + 1)- dimensional spacetime generically has no other symmetries.

The conformal Killing equations for the kinetic metric $\sigma_{i j}$ can be written in the form \begin{equation}
\sigma_{ij,k}{\bf Y}^k + \sigma_{i k} {\bf Y}^k_{,j} + \sigma_{k j} {\bf Y}^k_{,i} = 2 \psi (q^{\ell}) \sigma_{i j} \, , \label{CKV-eq}
\end{equation}
where $q^{\ell} = \{ N, A, Q, P, R \}, \, i,j,k,\ell = 1,2,3,4,5$. Now, we use the kinetic metric of the configuration space given in \eqref{km-fr} to look into the metric symmetry. The conformal Killing equations \eqref{CKV-eq} for this kinetic metric are obtained as
\begin{eqnarray}
& & f_{R} Y^3_{,A} = 0, \,\,   \nonumber \\& & f_{R} Y^4_{,N} + f_{RR} P Y^5_{,N} = 0, \,\, f_{R} Y^4_{,A} + f_{RR} P Y^5_{,A} = 0, \nonumber \\
& & f_{R} Y^1_{,P} + f_{RR} N Y^5_{,P} = 0, \,\, f_{R} Y^1_{,A} + f_{RR} N Y^5_{,A} = 0, \nonumber \\
& &  f_{RR} \left( P Y^1_{,A} + N Y^4_{,A} \right) = 0, \,\,  f_{RR} \left( P Y^1_{,R} + N Y^4_{,R} \right) = 0,  \nonumber \\
& &  f_{R} \left( \frac{P^3}{2 N} Y^3_{,N} + Y^4_{,Q} \right) + f_{RR} P Y^5_{,Q} = 0, \nonumber \\
& &  f_{R} \left( Y^1_{,Q} + \frac{P^3}{2 N} Y^3_{,P} \right) + f_{RR} N  Y^5_{,Q} = 0, \nonumber \\ & & f_{RR} \left( P Y^1_{,Q} + N Y^4_{,Q} \right) + f_{R} \frac{P^3}{2 N}  Y^3_{,R} = 0,  \label{keq} \\ & &  f_{R} \left( \frac{1}{A} Y^2 + Y^1_{,N} + Y^4_{,P} \right) + f_{RR} \left( Y^5 + N Y^5_{,N} + P Y^3_{,P} \right) = 2 \psi f_{R}, \nonumber \\
& &   f_{RR} \left( \frac{1}{A} Y^2 + \frac{1}{P} Y^4 + Y^1_{,N} + Y^5_{,R} + \frac{N}{P} Y^4_{,N} \right) + \frac{1}{P}f_R Y^4_{,R} + f_{RRR} Y^5 = 2 \psi f_{RR} \, , \nonumber \\
& &  f_{R} \left( -\frac{1}{N} Y^1 + \frac{1}{A} Y^2  + \frac{3}{P} Y^4 + 2 Y^3_{,Q} \right)  + f_{RR} Y^5 = 2 \psi f_{R} \, , \nonumber \\
& &  f_{RR} \left( \frac{1}{N} Y^1 + \frac{1}{A} Y^2 + Y^4_{,P} + Y^5_{,R} + \frac{P}{N} Y^1_{,P} \right) + \frac{1}{N} f_R Y^1_{,R} + f_{RRR} Y^5 = 2 \psi f_{RR} \nonumber
\end{eqnarray}
where $N, A, Q, P$ and $R$ are the configuration space variables, and ${\bf Y} = Y^1 \partial_N + Y^2 \partial_A + Y^3 \partial_Q + Y^4 \partial_P + Y^5 \partial_R $.

For any form of the function $f(R)$ under the condition $f_{RR} \neq 0$, we find \emph{six} KVs, which means $\psi =0$, as follows:
\begin{eqnarray}
& & {\bf Y}_1 = \partial_{Q}, \quad {\bf Y}_2 = A \partial_A + Q \partial_Q - P \partial_P, \nonumber \\ & & {\bf Y}_3 = N \partial_N - 2 A \partial_A + P \partial_P, \,\, {\bf Y}_4 = -A \partial_A + \frac{f_R}{f_{RR}} \partial_R ,  \nonumber \\ & & {\bf Y}_5 = N Q \partial_N +  \left( Q^2 + \frac{N^2}{P^2} \right) \partial_Q - P Q \partial_P, \label{KV} \\ & & {\bf Y}_6 = \ln ( N P) \left( N \partial_N + P \partial_P \right)  + A \ln \left( \frac{f_R^2}{N P} \right) \partial_A  - \frac{f_R \ln \left( N P f_R^2 \right) }{f_{RR}} \partial_R \, ,  \nonumber
\end{eqnarray}
with the non-vanishing Lie brackets
\begin{eqnarray}
& & \left[{\bf Y}_1,{\bf Y}_2 \right] = {\bf Y}_1, \quad \left[{\bf Y}_1,{\bf Y}_5 \right] = 2 {\bf Y}_2 + {\bf Y}_3, \nonumber \\ & &  \left[{\bf Y}_2,{\bf Y}_5 \right] = {\bf Y}_5, \quad \left[{\bf Y}_2,{\bf Y}_6 \right] = -{\bf Y}_3 + {\bf Y}_4, \label{alg-KV} \\ & & \left[{\bf Y}_3,{\bf Y}_6 \right] = 2 ( {\bf Y}_3 - {\bf Y}_4) \, , \,\, \left[{\bf Y}_4,{\bf Y}_6 \right] = -2 {\bf Y}_4. \nonumber
\end{eqnarray}
Thus, using the above KVs of the configuration space, it is seen that some of KVs are also Noether symmetries, that is, ${\bf X}_1 = {\bf Y}_1,$ $ {\bf X}_2 =  2 {\bf Y}_2 + {\bf Y}_3 $ and ${\bf X}_3 = {\bf Y}_5$. In the case of $\psi = constant$, it is found that there are \emph{seven} HVs of the configuration space which are
\begin{eqnarray}
& & {\bf Y}_1 = \partial_{Q}, \quad {\bf Y}_2^{HV} = Q \partial_Q - P \partial_P \quad {\rm with} \,\, \psi= -\frac{1}{2} \, , \nonumber \\ & & {\bf Y}_3^{HV} = N \partial_N  + P \partial_P \quad {\rm with} \,\, \psi= 1, \quad  {\bf Y}_4^{HV} = \frac{f_R}{f_{RR}} \partial_R \quad {\rm with} \,\, \psi= \frac{1}{2} \, ,  \nonumber \\ & & {\bf Y}_5 = N Q \partial_N +  \left( Q^2 + \frac{N^2}{P^2} \right) \partial_Q - P Q \partial_P, \label{HV} \\ & & {\bf Y}_6 = \ln ( N P) \left( N \partial_N + P \partial_P \right)  + A \ln \left( \frac{f_R^2}{N P} \right) \partial_A - \frac{f_R \ln \left( N P f_R^2 \right) }{f_{RR}} \partial_R \, ,  \nonumber \\ & & {\bf Y}_7^{HV} = A \partial_A  \quad {\rm with} \quad \psi= \frac{1}{2} \, ,  \nonumber
\end{eqnarray}
and the corresponding homothetic Lie algebra for the above HVs has the non-vanishing Lie brackets
\begin{eqnarray}
& & \left[{\bf Y}_1,{\bf Y}_2^{HV} \right] = {\bf Y}_1 \, , \,\,  \left[{\bf Y}_1,{\bf Y}_5 \right] = 2 {\bf Y}_2^{HV} + {\bf Y}_3^{HV} \, , \,\, \left[{\bf Y}_2^{HV},{\bf Y}_5 \right] = {\bf Y}_5 \, ,   \left[{\bf Y}_2^{HV},{\bf Y}_6 \right] = -{\bf Y}_3^{HV} + {\bf Y}_4^{HV} + {\bf Y}_7^{HV} \, , \nonumber \\ & & \left[{\bf Y}_3^{HV},{\bf Y}_6 \right] = 2 ( {\bf Y}_3^{HV} - {\bf Y}_4^{HV} - {\bf Y}_7^{HV} ) \, , \quad \left[{\bf Y}_4^{HV},{\bf Y}_6 \right] = 2(- {\bf Y}_4^{HV} + {\bf Y}_7^{HV} ) \, . \label{alg-HV}
\end{eqnarray}
If $\psi_{;\mu \nu} \neq 0$ which imposes that ${\bf Y}_i$'s are the CKV fields, then one can obtain from \eqref{keq} that the vector fields ${\bf Y}_1,...,{\bf Y}_5$ are the same as in \eqref{HV}, but ${\bf Y}_6$ and ${\bf Y}_7$ have the following form
\begin{eqnarray}
& & {\bf Y}_6^{CKV} = \ln ( N P) \left( N \partial_N + P \partial_P \right) + \Phi \ln \left( \frac{f_R^2}{N P} \right) \partial_A   - \frac{f_R \ln \left( N P f_R^2 \right) }{f_{RR}} \partial_R \,\, {\rm with} \,\, \psi= \frac{1}{2}\left( 1 - \frac{\Phi}{A} \right) \ln \left( \frac{N P}{f_R^2} \right) \, ,  \nonumber \\ & & {\bf Y}_7^{CKV} = \Phi \partial_A  \quad {\rm with} \quad \psi= \frac{\Phi}{2 A} \, , \label{CKV}
\end{eqnarray}
where the component $\Phi$ of ${\bf Y}_7$ is an arbitrary function of configuration space variables $N,A,Q,P$ and $R$. The corresponding conformal Killing algebra is 7-dimensional and the non-vanishing Lie brackets of this algebra are the same as \eqref{alg-HV} by changing ${\bf Y}_6$ and ${\bf Y}_7$ to ${\bf Y}_6^{CKV}$ and ${\bf Y}_7^{CKV}$, respectively,   if $\Phi = \Phi (A)$.

For the form of $f(R) = R - 2 \Lambda$ which requires that $f_{RR} = 0$, the equation \eqref{keq} with $\psi = 0$ yields \emph{ten} KVs such as
\begin{eqnarray}
& & {\bf Z}_1 = \partial_{Q}, \quad {\bf Z}_2 =  A \partial_A + Q \partial_Q - P \partial_P \, , \quad  {\bf Z}_3 = N \partial_N - A \partial_A + Q \partial_Q, \quad {\bf Z}_4 = P^2 \left( P \partial_P - 3 A \partial_A \right)  , \nonumber \\
& & {\bf Z}_5 = \frac{1}{N} \partial_N + \frac{A}{N^2} \partial_A \, , \quad
{\bf Z}_6 =  \frac{Q}{N} \partial_N  + \frac{Q A}{N^2} \partial_A  + \frac{1}{P^2} \partial_Q \, , \quad
{\bf Z}_7 =  Q {\bf Z}_4 - N^2 \partial_Q  \, , \nonumber \\
& &  {\bf Z}_8 = N Q \partial_N +  \left( Q^2 + \frac{N^2}{P^2} \right) \partial_Q - P Q \partial_P  \, ,  \quad {\bf Z}_9 = - Q {\bf Z}_7 + N^2 {\bf Z}_3 - 2 N^2 A \partial_A  \, ,  \label{KV-2} \\
& &   {\bf Z}_{10} = \frac{ Q^2}{2 N} \partial_N  + \frac{ ( P^2 Q^2 - N^2)}{2 N^2 P^2} A \partial_A + \frac{Q}{P^2} \partial_Q - \frac{1}{2 P} \partial_P \, . \nonumber
\end{eqnarray}
The non-vanishing Lie brackets of the above KVs are
\begin{eqnarray}
& & \left[{\bf Z}_1,{\bf Z}_2 \right] = {\bf Z}_1 \, , \quad \, \left[{\bf Z}_1,{\bf Z}_3 \right] = {\bf Z}_1 \, , \quad \,\, \left[{\bf Z}_1,{\bf Z}_6 \right] = {\bf Z}_5 \, , \nonumber \\ & & \left[{\bf Z}_1,{\bf Z}_7 \right] = {\bf Z}_4  \, ,  \quad \,  \left[{\bf Z}_1,{\bf Z}_8 \right] =  {\bf Z}_2 + {\bf Z}_3 \, , \, \left[{\bf Z}_1,{\bf Z}_9 \right] = -2 {\bf Z}_7 \, , \nonumber \\ & & \left[{\bf Z}_1,{\bf Z}_{10} \right] = {\bf Z}_6 \, , \,\,\,\,\, \left[{\bf Z}_2,{\bf Z}_4 \right] = - 2 {\bf Z}_4 \, , \left[{\bf Z}_2,{\bf Z}_6 \right] = {\bf Z}_6 \, , \nonumber \\ & &  \left[{\bf Z}_2,{\bf Z}_7 \right] = - {\bf Z}_7 \, , \,\,\, \left[{\bf Z}_2,{\bf Z}_8 \right] = {\bf Z}_8 \, ,  \quad \left[{\bf Z}_2,{\bf Z}_{10} \right] = 2 {\bf Z}_{10} \, , \nonumber \\ & & \left[{\bf Z}_3,{\bf Z}_5 \right] = - 2 {\bf Z}_5 \, , \left[{\bf Z}_3,{\bf Z}_6 \right] = -{\bf Z}_6 \, , \left[{\bf Z}_3,{\bf Z}_7 \right] = {\bf Z}_7 \, , \nonumber \\ & & \left[{\bf Z}_3,{\bf Z}_8 \right] = {\bf Z}_8 \, , \,\, \quad \left[{\bf Z}_3,{\bf Z}_9 \right] = 2 {\bf Z}_9 \, , \label{alg-KV-2} \\ & &  \left[{\bf Z}_4,{\bf Z}_6 \right] = - 2 {\bf Z}_1 \, , \left[{\bf Z}_4,{\bf Z}_8 \right] = 2 {\bf Z}_7 \, , \,  \left[{\bf Z}_4,{\bf Z}_{10} \right] = -2 {\bf Z}_2 \, ,  \nonumber  \\ & & \left[{\bf Z}_5,{\bf Z}_7 \right] = -2 {\bf Z}_1 \, , \, \left[{\bf Z}_5,{\bf Z}_8 \right] = 2 {\bf Z}_6 \, , \, \left[{\bf Z}_5,{\bf Z}_9 \right] = 4 {\bf Z}_3 \, ,   \nonumber \\ & & \left[{\bf Z}_6,{\bf Z}_7 \right] = - {\bf Z}_2 + {\bf Z}_3 \, , \left[{\bf Z}_6,{\bf Z}_8 \right] = 2 {\bf Z}_{10} \, , \left[{\bf Z}_6,{\bf Z}_9 \right] = 2 {\bf Z}_8 \, ,  \nonumber \\ & &  \left[{\bf Z}_7,{\bf Z}_8 \right] = -{\bf Z}_9 \, , \quad \left[{\bf Z}_7,{\bf Z}_{10} \right] = -{\bf Z}_8 \, . \nonumber
\end{eqnarray}
Furthermore, if $\psi$ is a constant, then there are \emph{eleven} HVs for this case which are ${\bf Z}_1, {\bf Z}_4,...,{\bf Z}_8$ and ${\bf Z}_{10}$ given in \eqref{KV-2} and the following ones
\begin{eqnarray}
& & {\bf Z}_2^{HV} =  Q \partial_Q - P \partial_P \quad {\rm with } \,\, \psi= -\frac{1}{2} \, , \nonumber \\ & &  {\bf Z}_3^{HV} = N \partial_N + Q \partial_Q \quad {\rm with} \,\, \psi= \frac{1}{2}  \, , \nonumber \\ & & {\bf Z}_9 = - Q {\bf Z}_7 + N^2 {\bf Z}_3^{HV} - 3 N^2 A \partial_A  \, ,   \label{HV-2} \\ & &  {\bf Z}_{11} = A \partial_A \quad {\rm with} \,\, \psi = \frac{1}{2} \, . \nonumber
\end{eqnarray}
The non-vanishing Lie brackets for those of HVs are almost the same with \eqref{alg-KV-2}, but the following ones are different:
\begin{eqnarray}
& & \left[{\bf Z}_4,{\bf Z}_{10} \right] = -2 \left( {\bf Z}_2^{HV} + {\bf Z}_{11} \right) \, , \quad \left[{\bf Z}_5,{\bf Z}_9 \right] = 4 ( {\bf Z}_3^{HV} - {\bf Z}_{11} ) \, , \quad \left[{\bf Z}_6,{\bf Z}_7 \right] = - {\bf Z}_2^{HV} + {\bf Z}_3^{HV} - 2 {\bf Z}_{11} \, . \label{alg-HV-2}
\end{eqnarray}
The \emph{eleven} CKV fields for the form of $f(R) = R - 2 \Lambda$ are obtained such that ${\bf Z}_1, {\bf Z}_2^{HV}, {\bf Z}_3^{HV}$ and ${\bf Z}_8$ are the same in \eqref{KV-2} and \eqref{HV-2}, and the remaining ones are
\begin{eqnarray}
& &  {\bf Z}_4^{CKV} = P^2 \left( P \partial_P - 3 \Psi \partial_A \right) \quad {\rm with} \,\, \psi = \frac{3}{2} \left( 1 - \frac{\Psi}{A} \right) P^2  \, , \nonumber \\
& & {\bf Z}_5^{CKV} = \frac{1}{N} \partial_N + \frac{\Psi}{N^2} \partial_A \qquad \quad {\rm with} \,\, \psi= \left( \frac{\Psi}{A} -1 \right) \frac{1}{N^2} \, , \nonumber \\
& & {\bf Z}_6^{CKV} =  \frac{Q}{N} \partial_N  + \frac{Q \Psi}{N^2} \partial_A  + \frac{1}{P^2} \partial_Q \quad {\rm with} \,\, \psi = \frac{1}{2} \left( \frac{\Psi}{A} -1 \right) \frac{Q}{N^2} \, , \nonumber \\
& & {\bf Z}_7^{CKV} = Q {\bf Z}_4^{CKV} - N^2 \partial_Q  \qquad \quad {\rm with} \quad \psi = \frac{3}{2} \left( 1 - \frac{\Psi}{A} \right) Q P^2 \, ,\label{CKV-2} \\
& & {\bf Z}_9^{CKV} = - Q {\bf Z}_7^{CKV} + N^2 {\bf Z}_3^{HV} - 3 N^2 \Psi \partial_A  \quad {\rm with} \,\, \psi = \frac{1}{2} \left( \frac{\Psi}{A} - 1 \right) ( P^2 Q^2 - N^2) \, ,  \nonumber \\
& & {\bf Z}_{10}^{CKV} = \frac{ Q^2}{2 N} \partial_N  + \frac{ ( P^2 Q^2 - N^2)}{2 N^2 P^2} \Psi \partial_A + \frac{Q}{P^2} \partial_Q  - \frac{1}{2 P} \partial_P \quad {\rm with} \,\, \psi = \frac{1}{4} \left( \frac{\Psi}{A}  - 1\right) \frac{ (P^2 Q^2 - N^2)}{N^2 P^2} \, , \nonumber \\
& &  {\bf Z}_{11}^{CKV} = \Psi \partial_A \quad {\rm with} \quad \psi = \frac{\Psi}{2 A} \, , \nonumber
\end{eqnarray}
where $\Psi = \Psi (N,A,Q,P,R)$. If $\Psi = \Psi (A)$, then the algebra of the CKVs is closed with the non-vanishing Lie brackets similar to ones as in \eqref{alg-KV-2} by changing some of  ${\bf Z}_i$'s to the ${\bf Z}_i^{CKV}$, and the different non-vanishing Lie brackets take the form
\begin{eqnarray}
& & \left[{\bf Z}_4^{CKV},{\bf Z}_{10}^{CKV} \right] = - 2 \left( {\bf Z}_2^{HV} + {\bf Z}_{11}^{CKV} \right) \, , \nonumber \\ & &  \left[{\bf Z}_5^{CKV},{\bf Z}_9^{CKV} \right] = 4 ( {\bf Z}_3^{HV} - {\bf Z}_{11}^{CKV} )\, ,  \label{alg-CKV-2} \\ & & \left[{\bf Z}_6^{CKV},{\bf Z}_7^{CKV} \right] = - {\bf Z}_2^{HV} + {\bf Z}_3^{HV} - 2 {\bf Z}_{11}^{CKV} \, . \nonumber
\end{eqnarray}
It is explicitly seen here that the CKVs reduces to the HVs if $\Psi = A$.

It can also be seen that the Noether symmetries  ${\bf X}_1 = \partial_Q, \quad {\bf X}_2 = N \partial_N + 2 \partial_Q - P \partial_P $ and  ${\bf X}_3 = N Q \partial_N + ( Q^2 + N^2 P^{-2} ) \partial_Q - P Q \partial_P $ are the KVs of the configuration space for any form of the function $f(R)$. Using the above symmetry equations \eqref{keq}, we found that the fourth Noether symmetry ${\bf X}_4 = \partial_r$ for any form $f(R)$ is \emph{not} a metric symmetry of the configuration space. The fifth Noether symmetry ${\bf X}_5$ given by \eqref{X5-r2} for $f(R)= f_0 R^n$ is a linear combination of the HVs obtained in \eqref{HV} such as
\begin{equation}
{\bf X}_5 = {\bf Y}_3^{HV} - \frac{4}{ 2n-1} {\bf Y}_4^{HV} - \frac{2}{2n-1} {\bf Y}_7^{HV}.
\end{equation}
This means that ${\bf X}_5$ for $f(R)= f_0 R^n$ is the HV of the configuration space. For the function $f(R) = R - 2\Lambda$, we observe that \emph{six} of seven Noether symmetries are the KVs of the configuration space.

\newpage


\end{document}